\documentclass[review]{elsarticle}

\usepackage{hyperref}  
\usepackage{xcolor}

\journal{Journal of New Astronomy}

\begin{document}

\begin{frontmatter}

\title{Energy rates due to Fe isotopes during presupernova evolution of massive stars}




\author{Jameel-Un Nabi}
\address{University of Wah, Quaid Avenue, Wah Cantt 47040, Punjab, Pakistan.}
\address{Faculty of Engineering Sciences, Ghulam Ishaq Khan Institute of Engineering Sciences and Technology, Topi, 23640, KP, Pakistan.}

\author{Asim Ullah}
\address{Faculty of Engineering Sciences, Ghulam Ishaq Khan Institute of Engineering Sciences and Technology, Topi, 23640, KP, Pakistan.}
\cortext[mycorrespondingauthor]{Corresponding author}
\ead{asimullah844@gmail.com}

\author{Majid Iqbal}
\address{University of Wah, Quaid Avenue, Wah Cantt 47040, Punjab, Pakistan.}

\begin{abstract}   
This work presents the microscopic calculation of energy rates ($\gamma$-ray heating  and (anti)neutrino cooling rates) due to weak decay of selected Fe Isotopes. The isotopes have astrophysical significance during the presupernova evolution of massive stars. The energy rates are calculated using the  pn-QRPA model  and compared with the independent particle model (IPM), large scale shell model (LSSM) {and recent shell model calculation (GXPF1J)}. The reported  (anti)neutrino cooling rates are smaller  by up to two orders of magnitude at low core temperature values than the IPM rates. The two calculations compare well at T = 30 GK. The comparison of cooling rates with the LSSM is interesting. The pn-QRPA cooling rates due to even-even Fe isotopes are smaller (up to 2 orders of magnitude). For the odd-A isotopes, the reported rates are bigger up to an order of magnitude.  {The pn-QRPA computed cooling rates are, up to 2 orders of magnitude, bigger when compared with the GXPF1J calculation.}  The $\gamma$-ray heating rates due to electron capture rates rise with the temperature and density values of the stellar core. On the other hand, the $\gamma$-ray heating due to $\beta$-decay increases with the core temperature values but decreases by orders of magnitude when the stellar core stiffens. { The pn-QRPA computed $\gamma$ heating rates are bigger (up to 3 orders of magnitude) at high temperatures and densities (for the case of $^{55,56}$Fe) when compared with the recent shell model results.}   Owing to the importance of energy rates, this study may contribute to a realistic simulation of presupernova evolution of massive stars.  
\end{abstract} 


\end{frontmatter}

\section{Introduction}
The nuclear matter within the core of a heavy-mass star is critical during the supernova explosion process and controls the evolutionary path of the massive stars~\cite{Bet79}. Although there have been enormous technological breakthroughs since Colgate \& White \cite{Col66} and Arnett \cite{Arn67} published their classic work on energy transport by (anti)neutrinos in non-rotating massive stars, the explosion mechanism of core-collapse supernovae is still not well understood.   Many physical inputs are required at the start of each stage of the simulation process, comprising of, but not limited to, core collapse, shock wave formation, stalling $\&$ revival, and shock propagation. Calculating the presupernova stellar structure with the most reliable physical data and inputs is highly desirable. \\ 
The reactions, mediated by the weak interaction, play crucial role during the evolutionary phases of massive stars \cite{Bet90,Lan03}. The conversion of proton into neutron via the weak interaction initiates the hydrogen burning in the stellar core of the massive star. The core later goes through He, C, Ne, O and Si burning phases. The core of the star evolves into Fe-peak nuclei following the Si burning phase. When the mass of the Fe core surpasses the Chandrasekhar mass limit ($\sim$ 1.5 M$_\odot$), the electron degeneracy pressure is no longer capable of supporting the gravitational contraction, and the core begins to collapse. The core-collapse dynamics is sensitive to the temporal variation of the lepton-to-baryon fraction (Y$_e$) and entropy of the core \cite{Bet79}. These quantities are primarily determined by the weak interaction (particularly electron capture (\textit{ec}) and $\beta$-decay (\textit{bd}) rates). The \textit{ec} reduces the Y$_e$, which in turn lowers the electron degeneracy pressure while the \textit{bd} does the opposite. The \textit{ec} and \textit{bd} reactions produce copious amount of neutrinos and anti-neutrinos, respectively. These fermions escape the star at densities $\rho$ $\leq$ 10$^{11}$ g/cm$^{3}$, channel energy and entropy away from the core and may expedite the collapse \cite{Heg01,Jan07}. These (anti)neutrinos are crucial for the dynamics of the stellar evolution. They provide information on neutronization caused by \textit{ec}, the infall phase, shock wave formation $\&$ propagation, and the cooling phase. Cooling rate is one of the critical parameters that significantly affect the stellar evolution. In the late evolutionary phases of heavy-mass stars, a change in the cooling rate can influence the time scale and configuration of the Fe core at the time of explosion \cite{Esp03}. The \textit{ec} rates and associated (anti)neutrino energy loss rates must also be taken into account when determining the equation of state. The neutrino energy loss rates can be used as essential inputs in multi-dimensional simulation study of the cooling proton-neutron stars. Depending on the initial mass and composition of the core, \textit{ec} to excited states and the resulting $\gamma$ emission can release up to ten times more heat, according to Gupta et al. \cite{Gup07}. The part of energy lost in a reaction caused by emitted neutrinos is significantly reduced. The capture of electrons on different nuclei, as well as the accompanying $\gamma$ heating, is temperature sensitive \cite{Gin15}. As a result, in addition to neutrino cooling rates, competing $\gamma$-ray heating rates for nuclei in the iron regime are useful.

The computation of stellar weak interaction rates including the associated  energy rates ($\gamma$-ray heating and (anti)neutrino cooling rates) during the late evolutionary stages of heavy-mass stars depend on reliable calculation of the ground and excited states Gamow-Teller response \cite{Bet79}.  During the initial stage of core-collapse, when the temperatures and densities are low (0.3 - 0.8 MeV and $\sim$ 10$^{10}$ g/cm$^3$, respectively), the Fermi energy and nuclear Q-value have comparable magnitudes. In such a scenario, the weak interaction rates depend more on the structure of the GT strength distributions. Once the chemical potential surpasses the Q-value at relatively high core densities, the weak interaction rates are controlled, more or less,  by the GT centroid and total GT strength values. Therefore, a detailed overview of the ground and excited states GT strength distributions is necessary for a reliable computation of the stellar weak interaction and associated energy rates. 
 
There have been numerous attempts in the past to calculate stellar weak interaction rates. Few noticeable mentions of weak rate calculations include Fuller et al. \cite{FFN}, Aufderheide et al. \cite{Auf94}, Nabi et al. \cite{Nab99},  Langanke et al. \cite{Lan00},  Pruet et al. \cite{Pru03} and Juodagalvis et al. \cite{Jug10}.   The \textit{ec} rates for various \textit{fp}-shell nuclei were recently calculated employing the pn-QRPA theory \cite{nab19}.   
In this study, we focus on the calculation of energy rates ($\gamma$-ray heating and (anti)neutrino cooling rates) due to weak rates on selected Fe isotopes ($^{55-58~ \& ~61-63}$Fe). { Earlier a study of neutrino cooling rates during presupernova evolution of massive stars for $^{54,55,56}$Fe was performed using the same model~\cite{Nab12}. In this follow-up project, we investigate a larger pool of Fe isotopes and also include the corresponding gamma heating rates. The Brink-Axel hypothesis~\cite{Bri58} was not used in the current calculation. This hypothesis assumes that the GT strength distributions for the parent excited states are the same as that of the ground state, only shifted by the energy of the excited state. For core temperatures exceeding 5 GK and densities in excess of 10$^7$ (g/cm$^3$), the ‘no-Brink’ and ‘Brink’ rates start to deviate, up to three orders of magnitude or more, for $sd$-shell nuclei~\cite{Nab22}. The calculated beta decay rates are affected more by invoking Brink-Axel hypothesis in the calculation as compared to the electron capture rates~\cite{Nab22}. For fp-shell nuclei, weak rates, based on the Brink-Axel hypothesis, start to deviate from the microscopically calculated rates at core temperatures exceeding 3 GK and densities beyond 10$^6$ (g/cm$^3$)~\cite{Fak23}. At high temperature and densities, the GT strength distributions of excited states are significantly different than those of the ground state~\cite{Nab09}. Therefore, the Brink-Axel hypothesis is not a good approximation to be used in stellar rate calculations, specially in high temperature-density environments.} These Fe isotopes are amongst the most relevant weak interaction nuclei according to the survey of Aufderheide et al. \cite{Auf94} and Heger et al. \cite{Heg01}.  According to a recent simulation study of presupernova evolution \cite{Nab21}, the selected Fe isotopes are among the top 30 important nuclei that significantly influences the dynamics of core-collapse. 

The paper is structured as follows: The theoretical framework used in the calculation of energy rates is concisely discussed in Section 2. The findings of the study are presented in Section~3. Summary and concluding remarks are 
given in the final section.

\section{Theoretical Formalism}
For the mean-field Hamiltonian, single-particle energies and wavefunctions were calculated employing the Nilsson model \cite{Nil55}. The residual interaction consisted of a schematic separable potential.
The pn-QRPA Hamiltonian chosen to calculate the energy rates associated with the weak interactions of the selected Fe isotopes was of the form:
\begin{equation} \label{H}
	H^{QRPA} = H^{sp} + V^{pp}_{GT} + V^{ph}_{GT} + V^{pair},
\end{equation}
where $H^{sp}$ is the single-particle Hamiltonian, $V_{GT}^{pp}$ ($V_{GT}^{ph}$) represents the particle-particle (particle-hole) GT
force and $V^{pair}$ stands for the pairing force calculated assuming BCS approximation. The advantage of using a simple pairing plus quadrupole Hamiltonian, with the incorporation of particle-hole and particle-particle GT forces of separable form, is that it transforms the diagonalization problem of the Hamiltonian matrix to solution of an algebraic equation of fourth order and greatly saves in computation time \cite{Mut92}. In addition, the choice of separable residual interaction permitted us to use a model space up to 7 major oscillator shell in our calculations allowing us to compute GT ground and excited states strength distributions for any arbitrary heavy nucleus.\\
{The $ph$ GT force ($V_{GT}^{ph}$) appearing in Eq.~\ref{H} was determined using:
\begin{equation}\label{ph}
	V^{ph}= +2\chi\sum^{1}_{\mu= -1}(-1)^{\mu}Y_{\mu}Y^{\dagger}_{-\mu},\\
\end{equation}
with
\begin{equation}\label{y}
	Y_{\mu}= \sum_{j_{p}m_{p}j_{n}m_{n}}<j_{p}m_{p}\mid
	t_- ~\sigma_{\mu}\mid
	j_{n}m_{n}>c^{\dagger}_{j_{p}m_{p}}c_{j_{n}m_{n}},
\end{equation}
whereas the $pp$ GT force ($V_{GT}^{pp}$) was computed using:
\begin{equation}\label{pp}
	V^{pp}= -2\kappa\sum^{1}_{\mu=
		-1}(-1)^{\mu}P^{\dagger}_{\mu}P_{-\mu},
\end{equation}
with
\begin{eqnarray}\label{p}
	P^{\dagger}_{\mu}= \sum_{j_{p}m_{p}j_{n}m_{n}}<j_{n}m_{n}\mid
	(t_- \sigma_{\mu})^{\dagger}\mid
	j_{p}m_{p}>\times
	(-1)^{l_{n}+j_{n}-m_{n}}c^{\dagger}_{j_{p}m_{p}}c^{\dagger}_{j_{n}-m_{n}},
\end{eqnarray}
In Eq.~(\ref{p}), the term $t_- \sigma_{\mu}$($t_+ \sigma_{\mu}$), represents the spherical part of the GT transformation operator that transforms neutron to proton (vice versa). The \textit{ph} and \textit{pp} forces have different signs revealing their opposite nature. The interaction strengths, particle-particle ($\kappa$) and particle-hole ($\chi$), were chosen in accordance with Ref.~\cite{Hom96}, based on a $1/A^{0.7}$ relationship.}
For computing the single-particle energies and wavefunctions, the Nilsson model  with the incorporation of nuclear deformation ($\beta$)
was employed. The oscillator constant for nucleons was computed using $\hbar\omega=\left(45A^{-1/3}-25A^{-2/3}\right)$ MeV \cite{Blo68} (same for protons and neutrons). The Nilsson-potential
parameters were adopted from Ref.~ \cite{ragnarson1984}. 
$Q$-values were computed from the recent mass compilation of Ref. \cite{Aud21}. The values of deformation parameter ($\beta$) were taken from Ref.~\cite{Mol16}. The pairing gaps for proton and neutron were computed using separation energies of proton (S$_p$) and neutron (S$_n$), respectively, as follows:
\begin{eqnarray}
	\bigtriangleup_{pp} =\frac{1}{4}(-1)^{Z+1}[S_p(Z+1, A+1)-2S_p(Z, A)+S_p(Z-1, A-1)]
\end{eqnarray}
\begin{eqnarray}
	\bigtriangleup_{nn} =\frac{1}{4}(-1)^{A-Z+1}[S_n(Z, A+1)- 2S_n(Z, A) + S_n(Z, A-1)]
\end{eqnarray} 
The (anti)neutrino energy loss rates in stellar environment can occur via four different types of weak-interaction
mediated channels: electron  (\textit{bd}) and positron emission (\textit{pe}), and continuum capture of electrons (\textit{ec}) and positrons (\textit{pc}). The (anti)neutrino energy loss rates from parent  ($\mathit{i}$) to the daughter nucleus ($\mathit{j}$) were computed using:
\begin{equation}
	\lambda_{ij}^{\nu(\bar{\nu})} =\left[\frac{ln2}{C}\right] \left[\phi_{ij}^{\nu(\bar{\nu})}(\rho, T, E_{f})\right] \left[B(F)_{ij}+(g_{A}/g_{V})^2 B(GT)_{ij} \right].
	\label{d_rate}
\end{equation}
We took $C$ = 6295$s$ and $g_A/g_V$ = -1.254 from Refs. \cite{Nak10}
and ~\cite{hardy09}, respectively. The reduced Fermi ($B(F)_{ij}$) and
GT transition probabilities ($B(GT)_{ij}$)  were calculated using the
following equations:
\begin{equation}
	B(F)_{ij} = \frac{1}{2J_{i}+1} \mid<j \parallel \sum_{k}t_{\pm}^{k}
	\parallel i> \mid ^{2},
\end{equation}
\begin{equation}\label{gt}
	B(GT)_{ij} = \frac{1}{2J_{i}+1} \mid <j
	\parallel \sum_{k}t_{\pm}^{k}\vec{\sigma}^{k} \parallel i> \mid ^{2},
\end{equation}
where the terms $J_i$, $\vec{\sigma}^k$ and $t_\pm^k$ stand for total spin (parent state), Pauli spin matrices and iso-spin raising and lowering operators, respectively. \\
The  $\phi_{ij}^{\nu(\bar{\nu})}$, appearing in Eq.~(\ref{d_rate}), are the phase space integrals over total energy and were computed using:
\begin{equation}\label{ps}
	\phi_{ij} = \int_{1}^{w_{m}} w (w_{m}-w)^{3}({w^{2}-1})^{\frac{1}{2}} F(\pm Z,w)
	(1-G_{\mp}) dw,
\end{equation}
for electron (upper signs) or positron (lower signs) emission, or by
\begin{equation}\label{pc}
	\phi_{ij} = \int_{w_{l}}^{\infty} w (w_{m}+w)^{3}({w^{2}-1})^{\frac{1}{2}} F(\pm Z,w) G_{\mp} dw,
\end{equation}
for continuum positron (lower signs) or electron (upper
signs) capture. The variables $w_{m}$, $w$ and $w_l$ appearing in the phase space integrals denote the total \textit{bd} energy, total energy of the electron (or positron) (rest mass + kinetic) and total capture threshold for electron (or positron) capture, respectively.  {The terms G$_{-}$ and G$_{+}$ denote the electron and positron distribution functions, respectively, given by;
\begin{equation}
	G_- = [exp(\frac{E-E_f}{kT})+1]^{-1},
\end{equation}
\begin{equation}
	G_+ = [exp(\frac{E+2+E_f}{kT})+1]^{-1}.
\end{equation}
where E = $(\omega-1)$ and E$_f$ represent the kinetic and the Fermi energy of the fermions excluding the electron rest mass}, respectively.
 The Fermi functions ($F(\pm Z, w)$) were calculated using the recipe given by Gove and Martin ~\cite{Gov71}. 
The total (anti)neutrino energy loss rate per unit time per nucleus was calculated using:
\begin{equation}\label{ecbd}
	\lambda^{\nu(\bar{\nu})} =\sum _{ij}P_{i} \lambda^{\nu(\bar{\nu})} _{ij}.
\end{equation}
where $\lambda_{ij}^{\nu}$ ($\lambda_{ij}^{\bar{\nu}}$) is the sum of \textit{ec} and \textit{pe} (\textit{pc} and \textit{bd}) rates for the transition from the $\mathit{i}^{th}$ state of the parent to the $\mathit{j}^{th}$ state of the daughter nucleus. All the initial and final state rates were summed until sufficient convergence was attained. {Because of the prevailing high temperatures in the stellar interior, there is a finite occupation probability   of parent excited states. Consequently, there is a significant contribution of these partial rates to the total weak decays. Under the assumption of thermal equilibrium, the probability of occupation of parent state, denoted by $P_{i}$ in Eq.~(\ref{ecbd}), was calculated using
\begin{equation}\label{pi}
	P_{i} = \frac {exp(-E_{i}/kT)}{\sum_{i=1}exp(-E_{i}/kT)}.
\end{equation}}
In a similar fashion, the total gamma ray heating rate per unit time for a given nucleus was calculated using the relation:
\begin{equation}\label{gr}
	\lambda^{\gamma} =\sum _{ij}P_{i} \lambda _{ij}E_j.
\end{equation}
Here $\lambda_{ij}$ is the sum of the \textit{ec} and \textit{pe} rates ($\lambda_{ij}$ = $\lambda_{ij}^{ec}$ + $\lambda_{ij}^{pe}$) or the sum of \textit{pc} and \textit{bd} rates ($\lambda_{ij}$ = $\lambda_{ij}^{pc}$ + $\lambda_{ij}^{bd}$) for the transition $i \rightarrow j$. E$_j$ denotes the energy of the daughter's excited state. For further insight and details of the formalism we refer to \cite{nab04}.

\section{Results and Discussion}
Only those Fe isotopes were selected for the current calculation which appeared in the top 30 list of most important weak-interaction nuclei, published recently by Nabi et al. \cite{Nab21}. The selected nuclei were ranked amongst top 30 most relevant nuclei on the basis of their largest contribution to the time rate of change of lepton fraction ($\dot{Y_e}$) during presupernova evolution (see Table~7 of \cite{Nab21}). Following assumptions were made in our calculation:\\
$\bullet$ The core temperature is sufficiently high to ionise the atoms. Electrons are in a continuous state and follow the Fermi-Dirac distribution.\\
$\bullet$ At high core temperatures ($kT >$ 1 MeV), positrons appear in our calculation through electron-positron pair creation.  It is further assumed that the positrons also follow the Fermi-Dirac distribution.\\
$\bullet$ All excited levels, with energy less than S$_n$ (or S$_p$), decay to the ground state via $\gamma$ transitions. Excited states, with energy  more than S$_n$ (or S$_p$),  emit neutrons (or protons).\\
$\bullet$ The reported $\gamma$-ray heating rates are either due to the sum of \textit{ec} and \textit{pe} or sum of \textit{pc} and \textit{bd} rates.\\
$\bullet$ The neutrino cooling rates are calculated taking into consideration the sum of \textit{ec} and \textit{pe} rates. The antineutrino cooling rates, on the other hand, are calculated taking into consideration the sum of \textit{pc} and \textit{bd} rates. For the considered density range, the (anti)neutrinos escape the  core without any interaction with the stellar matter.\\

{The reported (anti)neutrino cooling rates were compared with the previous calculations performed using the independent particle model (IPM)~\cite{FFN}, large-scale shell model (LSSM) results~\cite{Lan00} and a recent shell model calculation (employing the GXPF1J interaction) taking into account screening effects on electron capture rates~\cite{Mor20}. It is pertinent to note that the reported $\gamma$-ray heating rates are compared only with the recent shell model calculation which we henceforth refer to as GXPF1J throughout this text. IPM and LSSM calculations  did not compute the $\gamma$-ray heating rates.}

\begin{figure}
	\centering
	\includegraphics[width=0.9\textwidth]{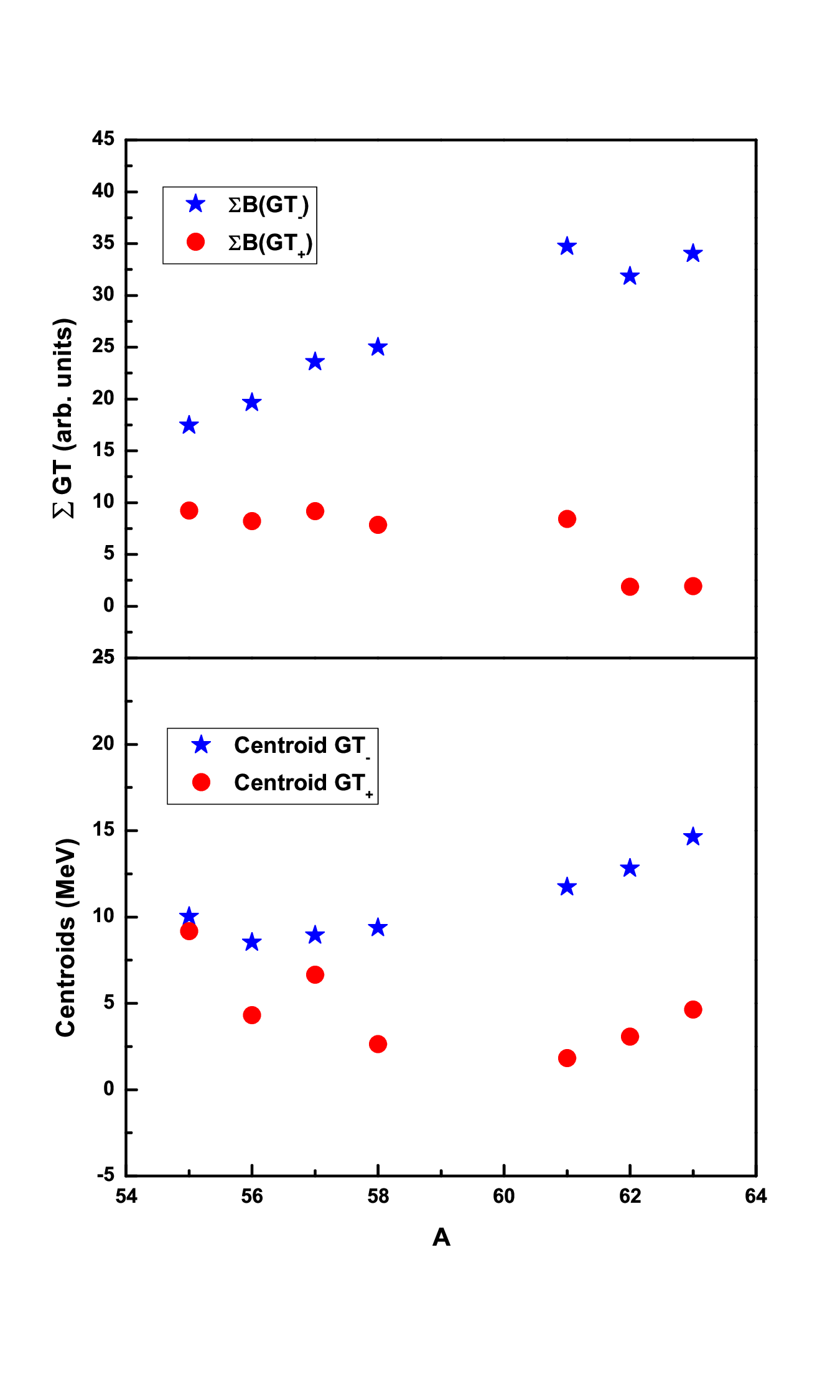}
	\vspace{-2cm}
	\caption{The pn-QRPA computed total GT strength (in arb. units) and GT centroids (MeV) for selected Fe isotopes.} \label{GT}
\end{figure}

The salient features of the ground-state GT strength distributions can be noted from Figure~\ref{GT}. The figure shows the calculated total GT strength (in both decay directions). Shown also are the computed GT centroid values. It is noted that Ikeda sum rule (ISR) is satisfied for all cases. 

\begin{figure}
	\centering
	\begin{tabular}{cc}
		\resizebox{0.6\hsize}{!}{\includegraphics*{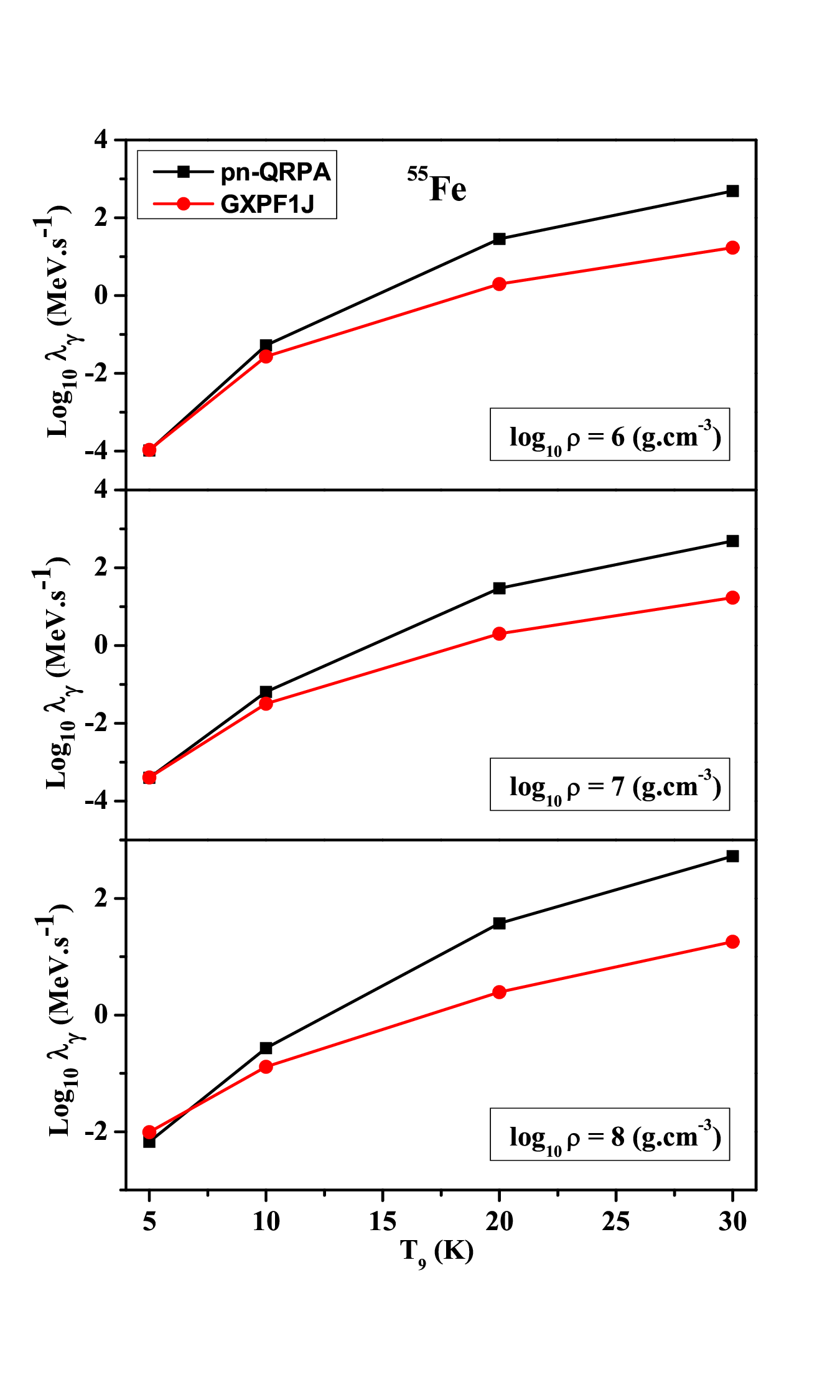}}&
		\resizebox{0.6\hsize}{!}{\includegraphics*{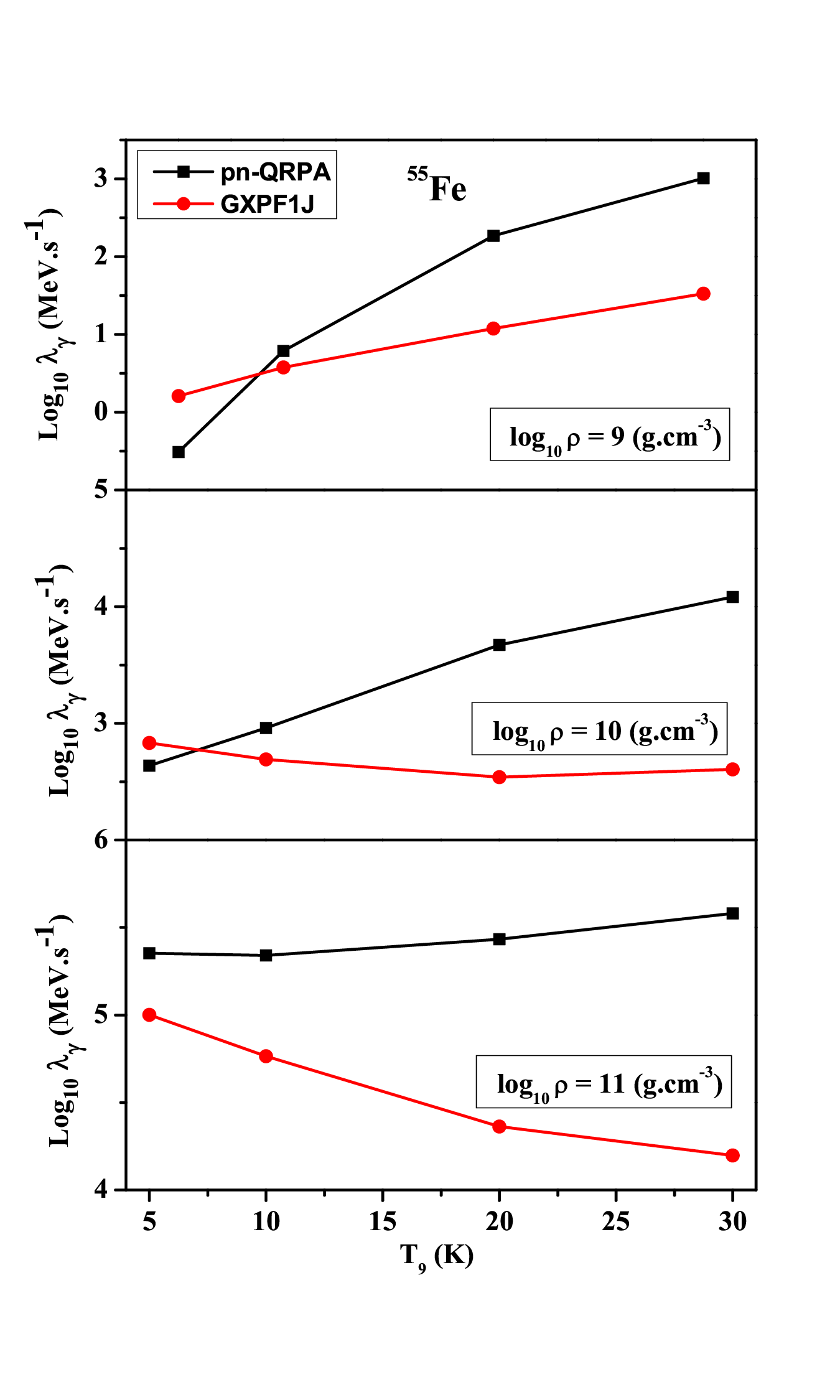}}\\
	\end{tabular}
	\vspace{-1cm}
	\caption{The pn-QRPA calculated $\gamma$-ray heating rates due to \textit{pe} and \textit{ec} rates on $^{55}$Fe, as a function of stellar temperature and density. The $\gamma$-ray heating rates are shown in log (to the base 10 scale) in units of MeV.s$^{-1}$.} \label{F1}
\end{figure}

\begin{figure}
	\centering
	\begin{tabular}{cc}
		\resizebox{0.6\hsize}{!}{\includegraphics*{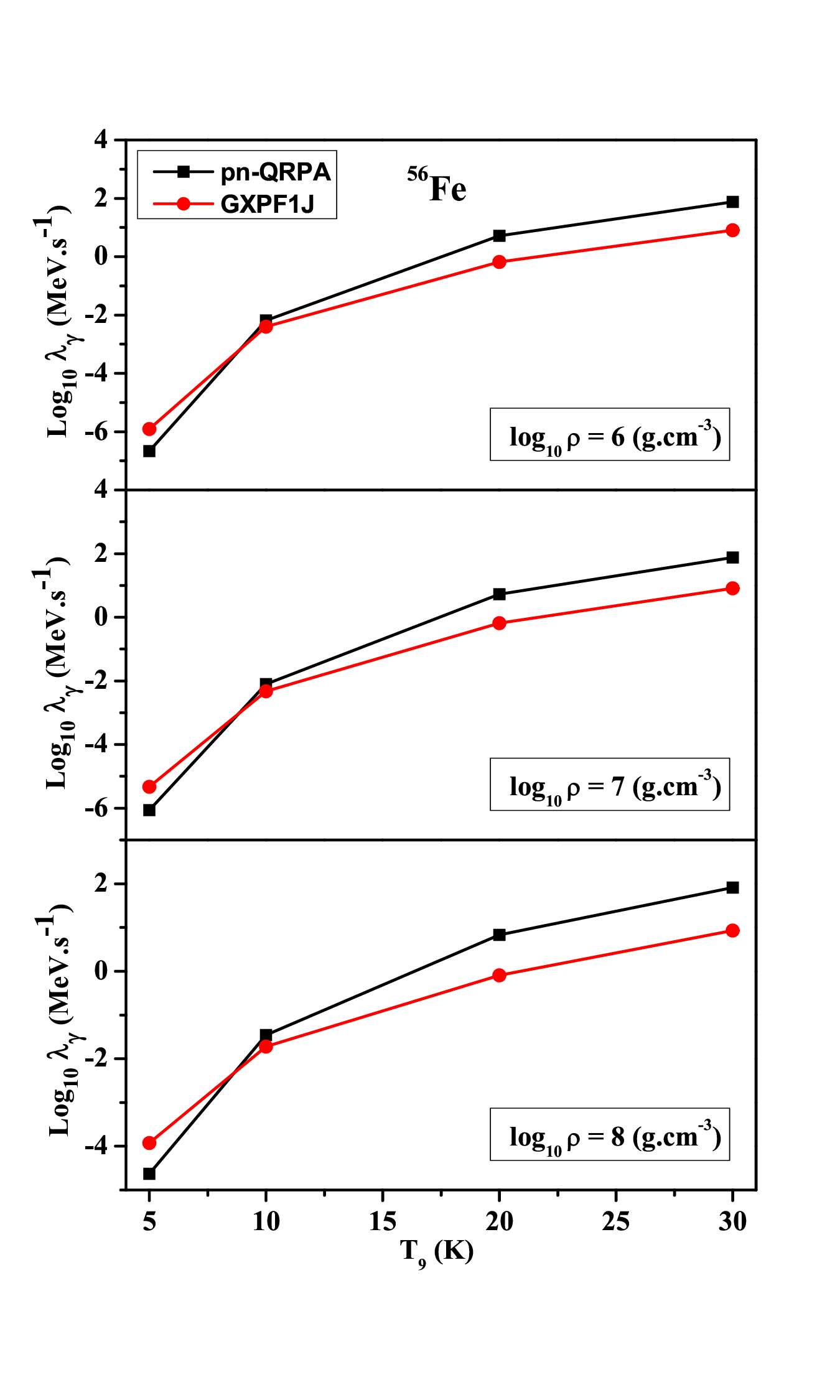}}&
		\resizebox{0.6\hsize}{!}{\includegraphics*{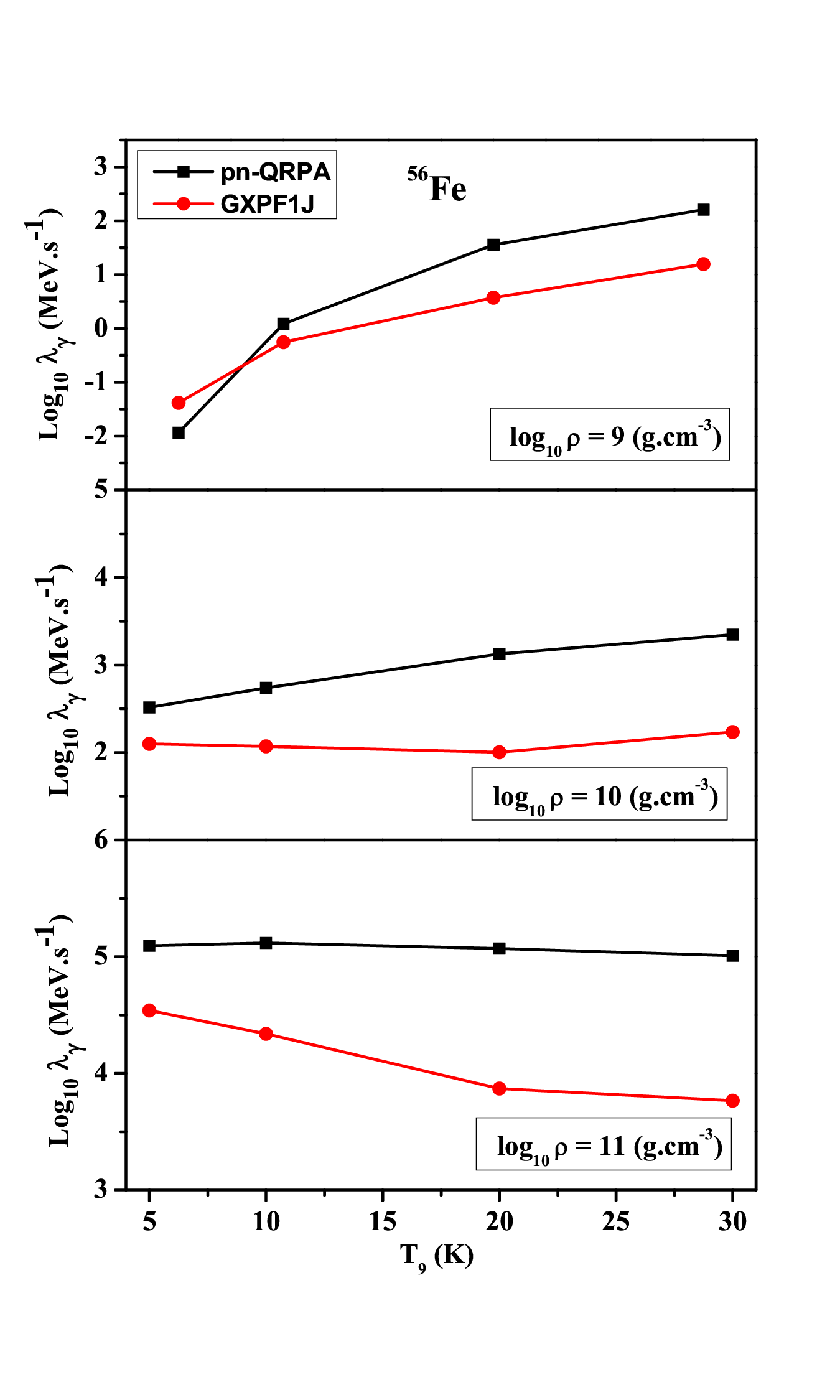}}\\
	\end{tabular}
	\vspace{-1cm}
	\caption{Same as Fig.~\ref{F1} but for $^{56}$Fe.} \label{F2}
\end{figure}

\begin{figure}
	\centering
	\begin{tabular}{cc}
		\resizebox{0.6\hsize}{!}{\includegraphics*{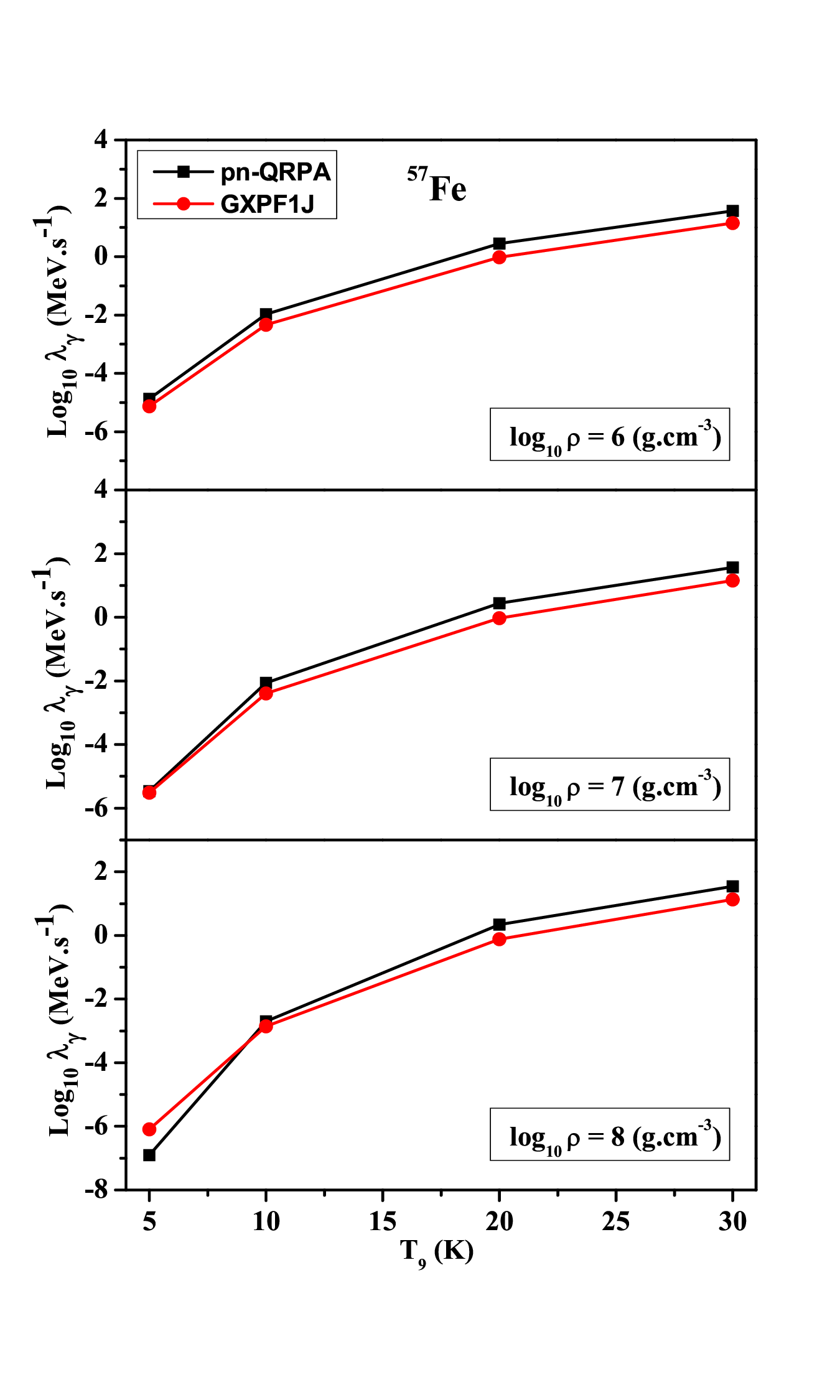}}&
		\resizebox{0.6\hsize}{!}{\includegraphics*{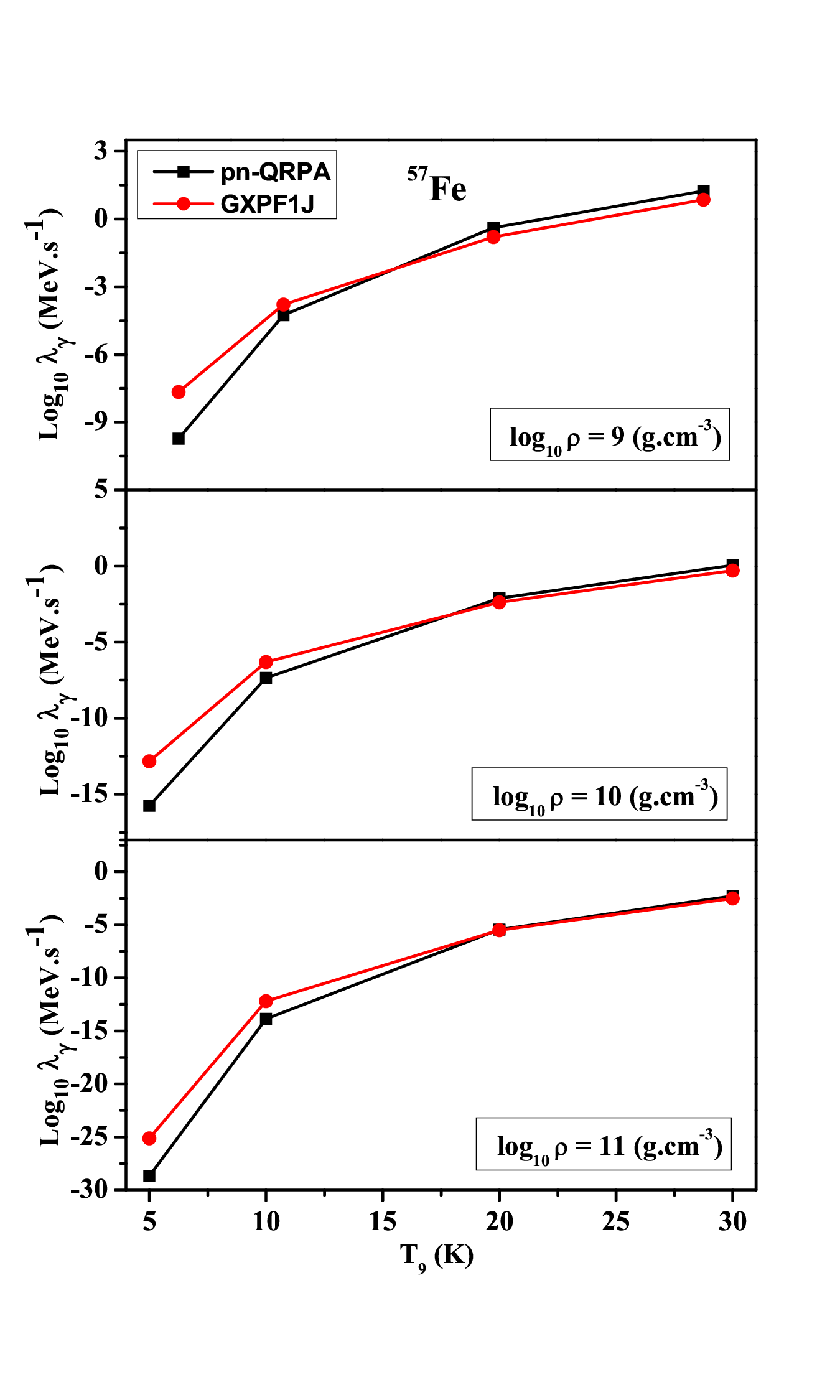}}\\
	\end{tabular}
	\vspace{-1cm}
	\caption{The pn-QRPA calculated $\gamma$-ray heating rates due to \textit{bd} and \textit{pc} rates on $^{57}$Fe in comparison with GXPF1J~\cite{Mor20}, as a function of stellar temperature and density. The $\gamma$-ray heating rates are shown in log (to the base 10 scale) in units of MeV.s$^{-1}$.} \label{F3}
\end{figure}

\begin{figure}
	\centering
\begin{tabular}{cc}
	\resizebox{0.6\hsize}{!}{\includegraphics*{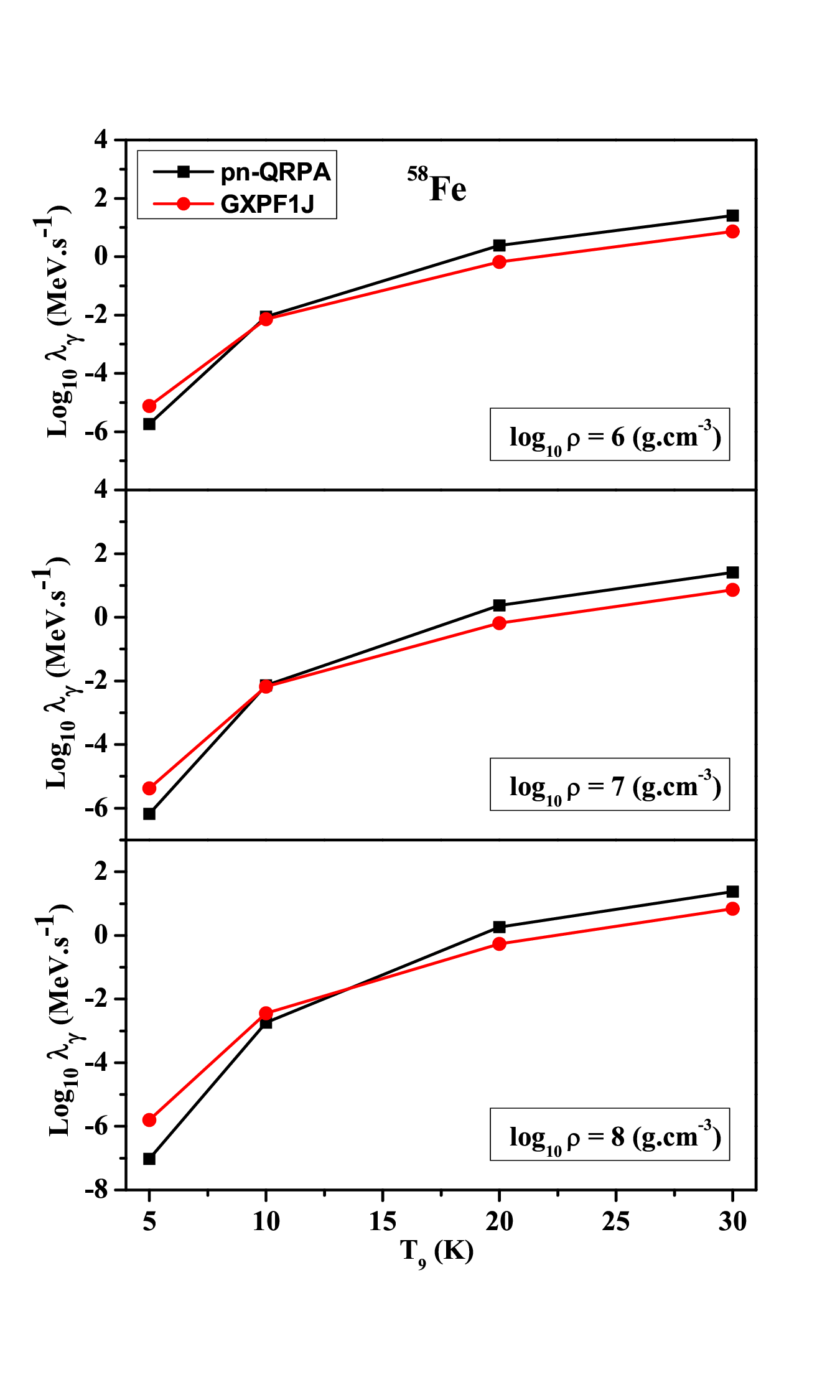}}&
	\resizebox{0.6\hsize}{!}{\includegraphics*{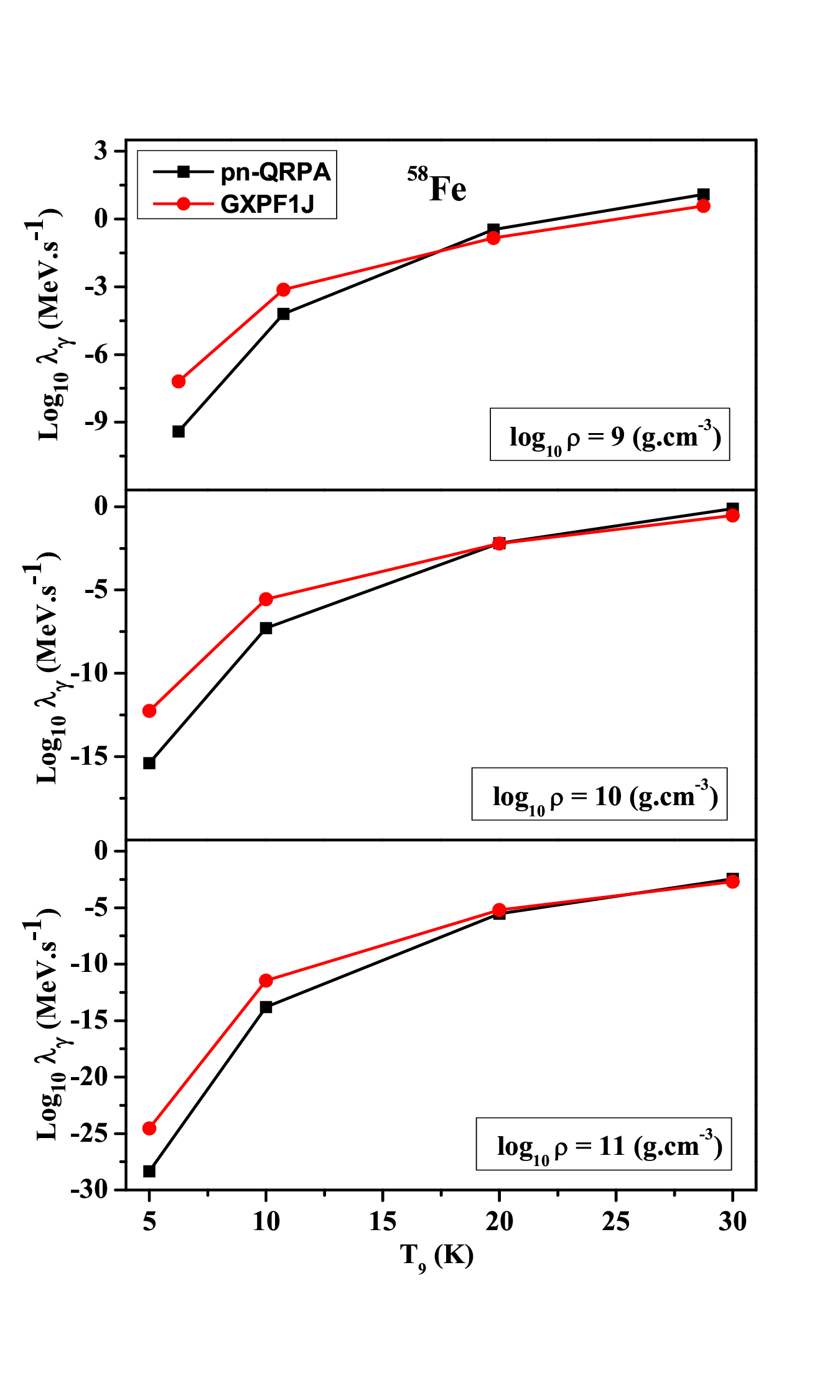}}\\
\end{tabular}
	\vspace{-1cm}
	\caption{Same as Fig.~\ref{F3} but for $^{58}$Fe.} \label{F4}
\end{figure}

\begin{figure}
	\centering
\begin{tabular}{cc}
	\resizebox{0.6\hsize}{!}{\includegraphics*{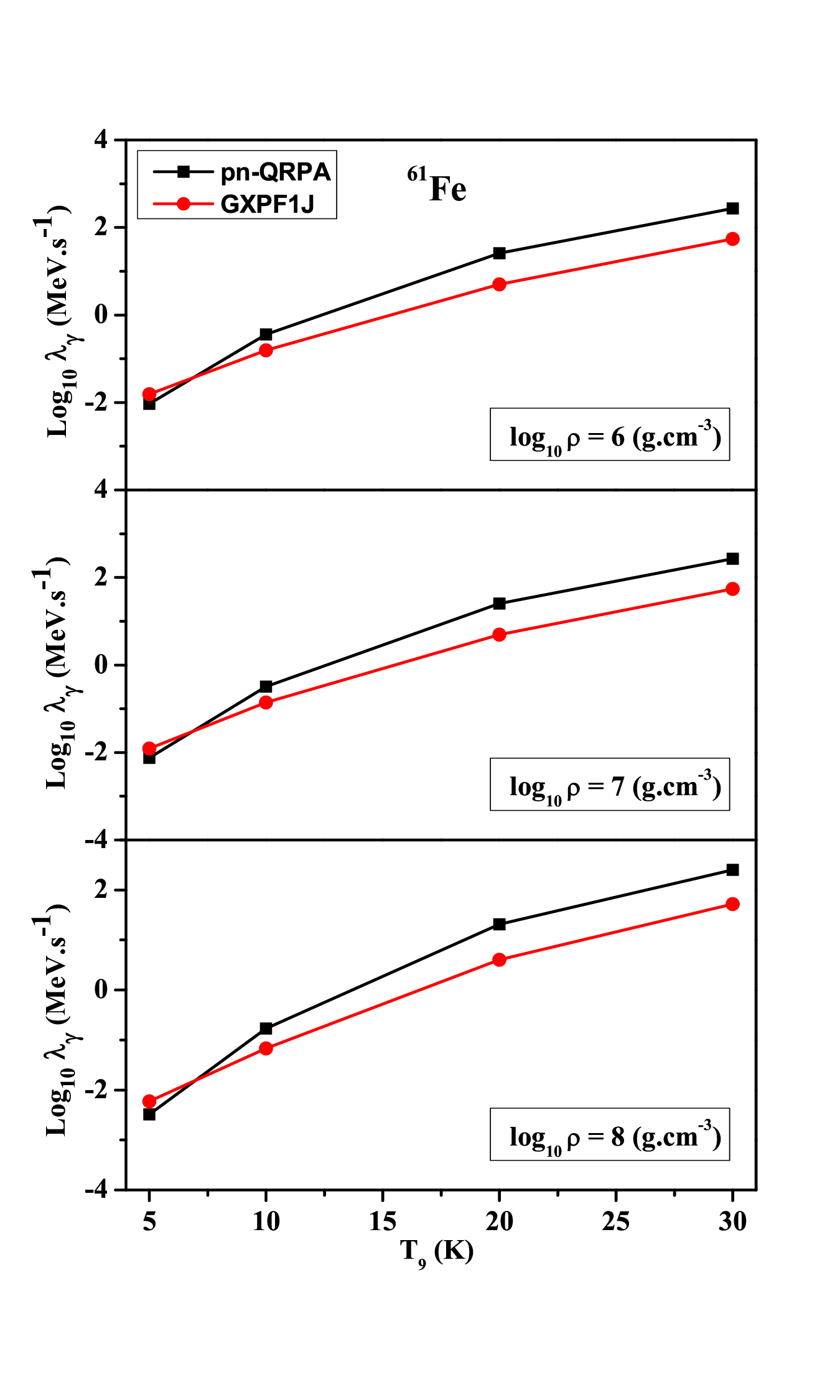}}&
	\resizebox{0.6\hsize}{!}{\includegraphics*{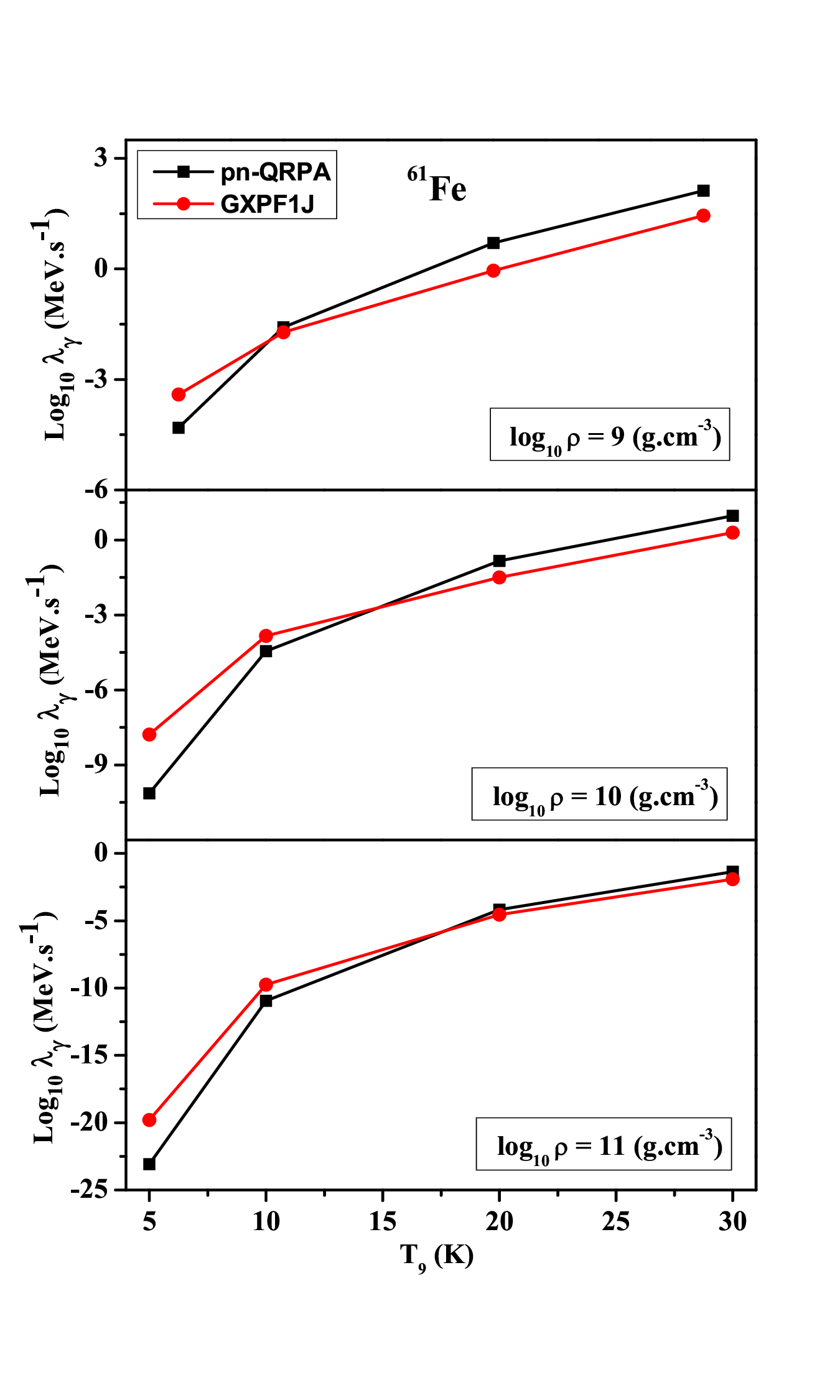}}\\
\end{tabular}
	\vspace{-1cm}
	\caption{Same as Fig.~\ref{F3} but for $^{61}$Fe.} \label{F5}
\end{figure}

\begin{figure}
	\centering
\begin{tabular}{cc}
	\resizebox{0.6\hsize}{!}{\includegraphics*{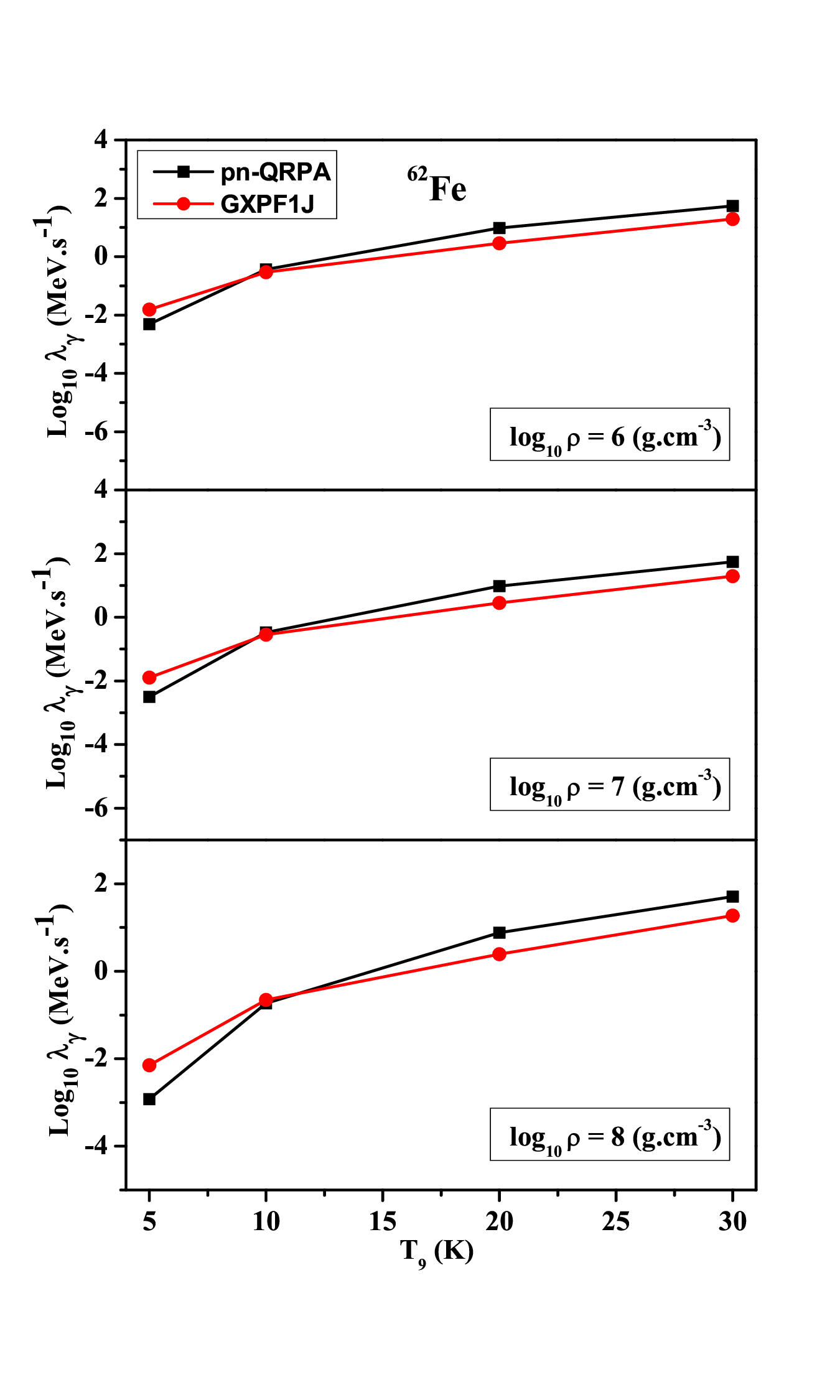}}&
	\resizebox{0.6\hsize}{!}{\includegraphics*{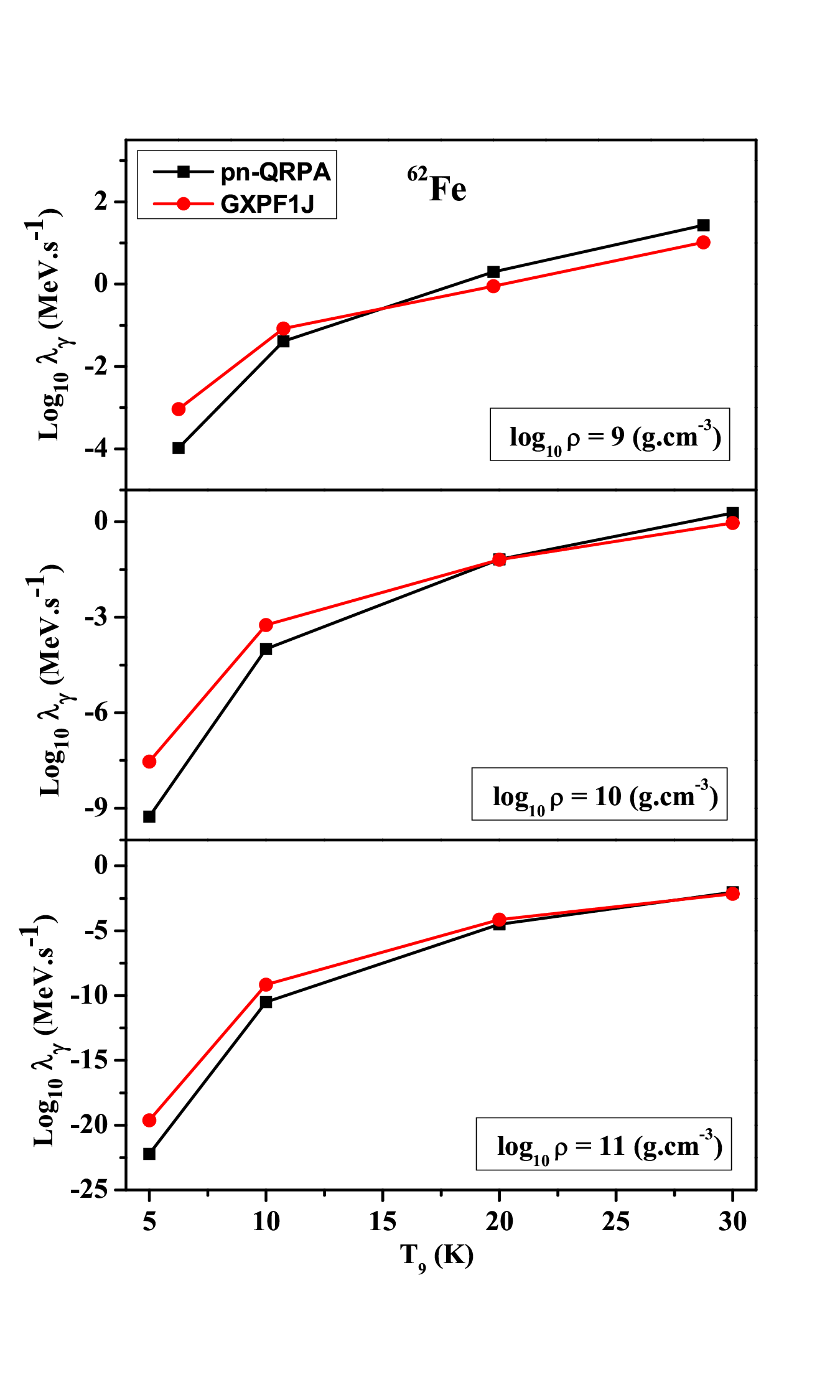}}\\
\end{tabular}
	\vspace{-1cm}
	\caption{Same as Fig.~\ref{F3} but for $^{62}$Fe.} \label{F6}
\end{figure}

Figures~(\ref{F1}~-~\ref{F2}) present the pn-QRPA calculated $\gamma$-ray heating rates due to \textit{pe} and \textit{ec} on $^{55}$Fe and $^{56}$Fe as a function of stellar temperature and core density. It is to be noted the $\gamma$-ray heating rates are shown in logarithmic scale in units of MeV/s. The $\gamma$-ray heating rates increase with rise in core temperature for reasons already stated above. It is noted that for core densities $\rho$ = [10 -- 10$^{6}$] (g/cm$^3$), the $\gamma$ heating rates do not change. At higher density values, the heating rates increase by orders of magnitude because of substantial increase in the Fermi energy thereby resulting in orders of magnitude enhancement in the \textit{ec} rates. 
Figures~(\ref{F3}~-~\ref{F6})  depict the $\gamma$-ray heating rates due to \textit{bd} and \textit{pc} on $^{57}$Fe, $^{58}$Fe, $^{61}$Fe \& $^{62}$Fe, respectively, as a function of stellar temperature and density values. Here the $\gamma$-ray heating rates follow a similar trend as discussed above. However, at high densities, the heating rates are suppressed by orders of magnitude. At higher densities, Pauli-blocking by the electrons in the final state poses a threshold for the \textit{bd}.
{Figures~(\ref{F1}~-~\ref{F6})  further depict a comparison of the pn-QRPA calculated heating rates with the GXPF1J results (which were reported only for stellar densities $\rho$ = [10$^{5}$ -- 10$^{11}$] (g/cm$^3$)). In comparison with GXPF1J, the pn-QRPA computed $\gamma$ heating rates are smaller at low T$_9$ values (up to 3 orders of magnitude). However, it is noted that the calculated rates are insignificant at such low temperatures and hardly bear any consequences for simulation studies. The comparison reverses at high core temperatures. For  $^{55,56}$Fe the pn-QRPA heating rates surpass the GXPF1J rates also at high density values. There are two main reasons for these differences. The GXPF1J rates incorporated quenching of the axial-vector coupling constant (g$_A^{eff}$/g$_{A}$ = 0.74) in their calculation. They further introduced screening effects in their computation which resulted in reduced capture rates. It is to be noted that $ec$ rates are sensitive to density values. As mentioned earlier the terrestrial decay mode of $^{55}$Fe is $ec$. The screening effects introduced by Ref.~\cite{Mor20} reduced their $ec$ rates (for $^{56}$Ni, the $ec$ rates with screening effects   were around  20\%–40\% smaller compared with those without these effects).  The gamma heating rates also include contributions from $\beta^{\pm}$ decays.} The current pn-QRPA model incorporates many configurations which translates into more parent excited states. At high temperatures, there is a finite occupation probability of these excited states and contribution of partial rates from these energy levels causes the orders of magnitude enhancement in our calculated rates.

Table~\ref{T1n} and Table~\ref{T2n} depict the comparison of pn-QRPA computed neutrino cooling rates with previous calculations, using the IPM, LSSM and {GXPF1J, for $^{55,56}$Fe} as a function of core temperature T$_9$ (in units of 10$^9$ K) and density (in units of g/cm$^3$).
Table~\ref{T1an} and Table~\ref{T2an} show a similar comparison for the anti-neutrino rates for $^{57,58}$Fe \& $^{61,62,63}$Fe under similar physical conditions. {It is again reminded that the GXPF1J calculation was not available for core densities $\rho$ $<$ 10$^5$ (g/cm$^3$). The shell model calculation was not performed for $^{63}$Fe.} The (anti)neutrino cooling rates are shown in units of MeV/s. The cooling rates increase with core temperature as more (anti)neutrinos are produced due to increase in weak rates (contribution from parent excited states enhances with increasing core temperatures).
The neutrino cooling rates increase as stellar core stiffens, owing to a substantial increment in Fermi energy. The anti-neutrino cooling rates are reduced at high density values due to Pauli-blocking by electrons in the final state
introducing a threshold for the \textit{bd} which ultimately has to be overcome by the thermal population of parent excited states.

In comparison with the IPM results, our calculated  cooling rates are generally smaller by up to two orders of magnitude at low core temperatures. The two calculations compare well  at high T$_9$ value of 30. The measured data obtained from different (n, p) and (p, n) experiments \cite{El-K94} had revealed the misplacement of the GT centroid in the parameterizations of IPM. Further, IPM calculation made use of the the Brink-Axel hypothesis \cite{Bri58} to compute excited states GT strength distributions. These are the most probable reasons for their bigger computed cooling rates at low temperatures. The pn-QRPA calculated  cooling rates for even-even cases ($^{56}$Fe, $^{58}$Fe \& $^{62}$Fe) are smaller by up to 2 orders of magnitude, when compared with the corresponding LSSM results. At  T$_9$ = 30, the reported rates are a few factor bigger than the LSSM numbers.  { For even-even nuclei, the pn-QRPA model calculates higher values of parent excited states (on the average 0.5 MeV higher). In the pn-QRPA model, excited states of an even-even nucleus are two-proton quasiparticle states and two-neutron quasiparticle states. When a nucleus has an odd nucleon, low-lying states are obtained by lifting the quasiparticle in the orbit of the smallest energy to higher lying orbits. States of odd-A Fe nuclei were expressed by three-neutron states or one-neutron two proton states, corresponding to excitation of a neutron or a proton.  Consequently, the calculated parent excited states are closer to measured levels for odd-A cases as compared to even-even cases. The probability of occupation of parent excited states becomes smaller with increasing energy values and translates to a smaller total cooling rate for even-even Fe isotopes (see Eq.~(\ref{ecbd})). In previous calculation~\cite{Nab12}, the authors replaced the higher computed pn-QRPA energy levels (for $^{55,56}$Fe) by measured levels and consequently reported bigger cooling rates as compared to present results.}   For odd-A cases, on the other hand, the LSSM results are generally smaller up to an order of magnitude, when compared with the pn-QRPA rates.  The LSSM placed the centroids of GT distributions of odd-A cases at much higher energies than IPM~\cite{Lan00}. Consequently the LSSM rates for odd-A cases are usually smaller. The rate convergence in LSSM approach depended on the number of Lanczos iteration and was limited up to 100 iterations only. This greatly affected their calculations especially at high temperatures.  It is to be noted that LSSM formalism, akin to the IPM calculation,  also made use of the Brink-Axel hypothesis for parent excited states in excess of a few MeVs. {The pn-QRPA computed (anti)neutrino cooling rates are generally bigger (up to 2 orders of magnitude) when compared with the GXPF1J calculation. Quenching and screening effects,  introduced by the shell model calculation, are the main sources for the reduced cooling rates.} Moreover, pn-QRPA model incorporates more configurations and includes many initial states which can be excited at high core temperatures.

\section{Summary and Conclusion} 
In this work we reported the calculation of energy rates ($\gamma$-ray heating and (anti)neutrino cooling rates) due to weak rates on selected Fe isotopes bearing astrophysical significance. The selected nuclei were included in the top 30 most relevant nuclei on the basis of their  contribution to the time rate of change of lepton fraction ($\dot{Y_e}$) during presupernova evolution according to a recent simulation study (see Table~7 of \cite{Nab21}). Feature of our project was the microscopic calculation of ground and excited-states GT strength distributions without assuming the Brink-Axel hypothesis \cite{Bri58}.  

The $\gamma$ heating rates do not change for the density range $\rho$ = [10 -- 10$^{6}$] (g/cm$^3$).  {The heating rates were compared with the  recent shell model calculation \cite{Mor20}. In comparison with GXPF1J results, the pn-QRPA computed $\gamma$ heating rates were found bigger (up to 3 orders of magnitude) at high temperatures and densities ($^{55,56}$Fe).} 

The computed (anti)neutrino cooling rates were compared with the results of IPM \cite{FFN}, LSSM \cite{Lan00} {and GXPF1J interaction}. In comparison with the IPM and LSSM calculations, the reported cooling rates were found smaller by up to two orders of magnitude at low T$_9$ values. Only for the case of odd-A Fe isotopes, the LSSM cooling rates were found smaller, up to an order of magnitude, when compared with the reported rates. {The pn-QRPA computed (anti)neutrino cooling rates were, up to 2 orders of magnitude, bigger when compared with the GXPF1J calculation.} The reported energy rates may prove useful for a realistic simulation of presupernova evolution of massive stars.

\section*{Acknowledgements}
J.-U. Nabi would like to acknowledge the support of the Higher Education Commission Pakistan through project
20-15394/NRPU/R\&D/HEC/2021.

\begin{table*}[]
	\scriptsize\caption{Comparison of the pn-QRPA calculated neutrino cooling rates with IPM~\cite{FFN}, LSSM~\cite{Lan00} and GXPF1J~\cite{Mor20} results for $^{55}$Fe and $^{56}$Fe. }
	\label{T1n}
	\centering
	\scalebox{.70}{
		\begin{tabular}{c|c|ccc|ccc|cccc|}
			\hline
			\multicolumn{6}{l}{} & \multicolumn{1}{l}{$\lambda^{\nu}$} (MeV.$s^{-1}$)& \multicolumn{5}{l}{}        \\
			\cline{1-12} \multicolumn{1}{l}{} &  & \multicolumn{3}{c}{$\rho$$\it Y_{e}$ = $10^1$ (g.cm$^{-3}$) }        & \multicolumn{3}{c}{$\rho$$\it Y_{e}$ = $10^3$ (g.cm$^{-3}$) }& \multicolumn{4}{c}{$\rho$$\it Y_{e}$ = $10^{5}$ (g.cm$^{-3}$)}  \\
			\cline{3-12} \multicolumn{1}{l}{Nuclei} &       $T_9$ (GK) & $\lambda_{pn-QRPA}^{\nu}$& $\lambda_{IPM}^{\nu}$    & $\lambda_{LSSM}^{\nu}$      &$\lambda_{pn-QRPA}^{\nu}$& $\lambda_{IPM}^{\nu}$    & $\lambda_{LSSM}^{\nu}$       & $\lambda_{pn-QRPA}^{\nu}$& $\lambda_{IPM}^{\nu}$    & $\lambda_{LSSM}^{\nu}$&$\lambda_{GXPF1J}^{\nu}$  \\
			\hline\\
			$^{55}$Fe & 1  & 4.63E-09 & 1.72E-10 & 1.67E-10 & 7.83E-09 & 2.90E-10 & 2.83E-10 & 5.19E-07 & 1.97E-08 & 1.92E-08 & 1.97E-08             \\
			& 2  & 3.34E-06 & 1.41E-07 & 6.27E-08 & 3.37E-06 & 1.42E-07 & 6.32E-08 & 7.00E-06 & 2.91E-07 & 1.30E-07 & 1.36E-07             \\
			& 3  & 6.71E-05 & 8.81E-06 & 2.17E-06 & 6.73E-05 & 8.81E-06 & 2.17E-06 & 7.71E-05 & 1.00E-05 & 2.47E-06 & 2.88E-06             \\
			& 5  & 2.04E-03 & 1.32E-03 & 1.87E-04 & 2.04E-03 & 1.32E-03 & 1.87E-04 & 2.08E-03 & 1.35E-03 & 1.91E-04 & 2.20E-04             \\
			& 10 & 1.97E-01 & 6.27E-01 & 8.73E-02 & 1.98E-01 & 6.27E-01 & 8.73E-02 & 1.98E-01 & 6.28E-01 & 8.75E-02 & 5.83E-02             \\
			& 30 & 3.95E+03 & 2.20E+03 & 4.36E+02 & 3.96E+03 & 2.20E+03 & 4.36E+02 & 3.96E+03 & 2.20E+03 & 4.36E+02 & 1.12E+02             \\
			&    &     &      &     &      &      &     &     &     &     &  \\
			$^{56}$Fe & 1  & 5.97E-29 & 1.85E-26 & 5.66E-27 & 1.01E-28 & 3.11E-26 & 9.57E-27 & 6.95E-27 & 2.21E-24 & 6.81E-25 & 2.49E-24             \\
			& 2  & 3.31E-16 & 3.30E-14 & 8.55E-15 & 3.34E-16 & 3.33E-14 & 8.63E-15 & 6.97E-16 & 6.95E-14 & 1.80E-14 & 3.92E-14             \\
			& 3  & 1.86E-11 & 1.03E-09 & 2.34E-10 & 1.87E-11 & 1.03E-09 & 2.35E-10 & 2.14E-11 & 1.18E-09 & 2.69E-10 & 5.24E-10             \\
			& 5  & 4.88E-07 & 9.35E-06 & 2.28E-06 & 4.88E-07 & 9.35E-06 & 2.28E-06 & 4.98E-07 & 9.55E-06 & 2.33E-06 & 4.22E-06             \\
			& 10 & 1.18E-02 & 3.09E-02 & 1.61E-02 & 1.18E-02 & 3.09E-02 & 1.61E-02 & 1.19E-02 & 3.10E-02 & 1.62E-02 & 1.77E-02             \\
			& 30 & 4.19E+02 & 7.03E+02 & 2.85E+02 & 4.20E+02 & 7.03E+02 & 2.85E+02 & 4.20E+02 & 7.03E+02 & 2.85E+02 & 1.50E+02           \\
			\hline
	\end{tabular}}
\end{table*}
\begin{table*}[]
	\scriptsize\caption{Same as Table~\ref{T1n} but for higher core densities.}
	\label{T2n}
	\centering
	\scalebox{.70}{
		\begin{tabular}{c|c|cccc|cccc|cccc|}
			\hline
			\multicolumn{7}{l}{} & \multicolumn{1}{l}{$\lambda^{\nu}$ (MeV.$s^{-1}$)} & \multicolumn{6}{l}{}        \\
			\cline{1-14} \multicolumn{1}{l}{} &  & \multicolumn{4}{c}{$\rho$$\it Y_{e}$ = $10^7$ (g.cm$^{-3}$)}        & \multicolumn{4}{c}{$\rho$$\it Y_{e}$ = $10^9$ (g.cm$^{-3}$)}& \multicolumn{4}{c}{$\rho$$\it Y_{e}$ = $10^{11}$ (g.cm$^{-3}$) }         \\
			\cline{3-14} \multicolumn{1}{l}{Nuclei} &       $T_9$ (GK) & $\lambda_{pn-QRPA}^{\nu}$& $\lambda_{IPM}^{\nu}$    & $\lambda_{LSSM}^{\nu}$  &$\lambda_{GXPF1J}^{\nu}$    &$\lambda_{pn-QRPA}^{\nu}$& $\lambda_{IPM}^{\nu}$    & $\lambda_{LSSM}^{\nu}$  &$\lambda_{GXPF1J}^{\nu}$     & $\lambda_{pn-QRPA}^{\nu}$& $\lambda_{IPM}^{\nu}$    & $\lambda_{LSSM}^{\nu}$&$\lambda_{GXPF1J}^{\nu}$  \\
			\hline\\
			$^{55}$Fe & 1  & 1.49E-04 & 1.03E-05 & 1.02E-05 & 8.95E-06             & 8.17E-01 & 1.79E+01 & 6.08E-01 & 6.32E-01             & 4.10E+05 & 1.92E+06 & 3.85E+05 & 2.88E+05             \\
			& 2  & 1.08E-03 & 3.99E-05 & 2.14E-05 & 1.98E-05             & 4.05E+00 & 1.96E+01 & 7.18E-01 & 7.36E-01             & 5.57E+05 & 1.92E+06 & 3.87E+05 & 2.90E+05             \\
			& 3  & 2.70E-03 & 3.21E-04 & 8.04E-05 & 8.49E-05             & 6.67E+00 & 2.29E+01 & 9.59E-01 & 9.53E-01             & 6.58E+05 & 1.93E+06 & 3.92E+05 & 2.90E+05             \\
			& 5  & 9.86E-03 & 6.40E-03 & 8.26E-04 & 9.18E-04             & 1.03E+01 & 3.55E+01 & 1.94E+00 & 1.73E+00             & 7.62E+05 & 1.95E+06 & 4.04E+05 & 2.74E+05             \\
			& 10 & 2.48E-01 & 7.83E-01 & 1.07E-01 & 6.90E-02             & 2.58E+01 & 1.18E+02 & 1.15E+01 & 5.74E+00             & 1.04E+06 & 2.04E+06 & 4.32E+05 & 1.56E+05             \\
			& 30 & 4.00E+03 & 2.22E+03 & 4.40E+02 & 1.12E+02             & 8.43E+03 & 4.65E+03 & 9.18E+02 & 2.17E+02             & 8.53E+06 & 3.08E+06 & 7.03E+05 & 8.85E+04             \\
			&    & &    &     & &     &    &      & &      &     &    &\\
			$^{56}$Fe & 1  & 2.92E-23 & 1.69E-20 & 5.45E-21 & 1.21E-20             & 2.82E-04 & 1.72E-03 & 6.34E-03 & 2.73E-03             & 4.15E+05 & 6.85E+05 & 2.27E+05 & 1.77E+05             \\
			& 2  & 1.64E-13 & 2.17E-11 & 5.71E-12 & 9.73E-12             & 6.85E-04 & 6.76E-03 & 1.18E-02 & 7.28E-03             & 4.16E+05 & 6.90E+05 & 2.29E+05 & 1.78E+05             \\
			& 3  & 8.18E-10 & 5.19E-08 & 1.17E-08 & 1.94E-08             & 1.78E-03 & 3.18E-02 & 2.88E-02 & 2.17E-02             & 4.18E+05 & 7.00E+05 & 2.33E+05 & 1.79E+05             \\
			& 5  & 2.37E-06 & 4.68E-05 & 1.12E-05 & 1.86E-05             & 1.50E-02 & 2.82E-01 & 1.45E-01 & 1.27E-01             & 4.22E+05 & 7.24E+05 & 2.45E+05 & 1.79E+05             \\
			& 10 & 1.48E-02 & 3.86E-02 & 2.01E-02 & 2.11E-02             & 1.91E+00 & 5.32E+00 & 2.66E+00 & 2.00E+00             & 4.54E+05 & 7.89E+05 & 2.61E+05 & 1.03E+05             \\
			& 30 & 4.24E+02 & 7.08E+02 & 2.87E+02 & 1.50E+02             & 8.95E+02 & 1.47E+03 & 6.00E+02 & 2.91E+02             & 9.62E+05 & 1.42E+06 & 4.69E+05 & 9.93E+04   \\
			\hline
	\end{tabular}}
\end{table*}
\begin{table*}[]
	\scriptsize\caption{Comparison of the pn-QRPA calculated anti-neutrino cooling rates with IPM~\cite{FFN}, LSSM~\cite{Lan00} and GXPF1J~\cite{Mor20} results for selected Fe isotopes.}
	\label{T1an}
	\centering
	\scalebox{.70}{
		\begin{tabular}{c|c|ccc|ccc|cccc|}
			\hline
			\multicolumn{6}{l}{} & \multicolumn{1}{l}{$\lambda^{\bar{\nu}}$ (MeV.$s^{-1}$)} & \multicolumn{5}{l}{}        \\
			\cline{1-12} \multicolumn{1}{l}{} &  & \multicolumn{3}{c}{$\rho$$\it Y_{e}$ = $10^1$ (g.cm$^{-3}$)}        & \multicolumn{3}{c}{$\rho$$\it Y_{e}$ = $10^3$ (g.cm$^{-3}$)}& \multicolumn{4}{c}{$\rho$$\it Y_{e}$ = $10^{5}$ (g.cm$^{-3}$)}         \\
			\cline{3-12} \multicolumn{1}{l}{Nuclei} &       $T_9$ (GK) & $\lambda_{pn-QRPA}^{\bar{\nu}}$& $\lambda_{IPM}^{\bar{\nu}}$    & $\lambda_{LSSM}^{\bar{\nu}}$      &$\lambda_{pn-QRPA}^{\bar{\nu}}$& $\lambda_{IPM}^{\bar{\nu}}$    & $\lambda_{LSSM}^{\bar{\nu}}$       & $\lambda_{pn-QRPA}^{\bar{\nu}}$& $\lambda_{IPM}^{\bar{\nu}}$    & $\lambda_{LSSM}^{\bar{\nu}}$  &$\lambda_{GXPF1J}^{\bar{\nu}}$\\
			\hline\\
		$^{57}$Fe & 1  & 3.66E-12 & 6.37E-12 & 4.22E-12 & 2.17E-12 & 4.37E-12 & 2.51E-12 & 4.39E-14 & 1.33E-12 & 4.72E-14 & 1.42E-14             \\
		& 2  & 2.23E-08 & 1.34E-08 & 2.62E-09 & 2.22E-08 & 1.33E-08 & 2.60E-09 & 1.15E-08 & 8.85E-09 & 1.43E-09 & 7.76E-10             \\
		& 3  & 9.89E-07 & 2.11E-06 & 1.34E-07 & 9.89E-07 & 2.11E-06 & 1.34E-07 & 8.79E-07 & 2.02E-06 & 1.26E-07 & 1.03E-07             \\
		& 5  & 7.03E-05 & 4.91E-04 & 3.85E-05 & 7.05E-05 & 4.91E-04 & 3.85E-05 & 6.92E-05 & 4.89E-04 & 3.84E-05 & 3.30E-05             \\
		& 10 & 1.90E-02 & 1.40E-01 & 1.65E-02 & 1.91E-02 & 1.40E-01 & 1.65E-02 & 1.90E-02 & 1.40E-01 & 1.65E-02 & 1.27E-02             \\
		& 30 & 1.04E+03 & 6.04E+02 & 4.02E+01 & 1.04E+03 & 6.04E+02 & 4.02E+01 & 1.04E+03 & 6.04E+02 & 4.02E+01 & 2.96E+01             \\
		&&&&&&&&&&&\\
$^{58}$Fe&1  & 1.54E-19 & 1.55E-16 & 9.62E-18 & 9.20E-20 & 9.18E-17 & 5.70E-18 & 3.73E-21 & 1.70E-18 & 8.95E-20 & 1.00E-19             \\
		& 2  & 7.10E-12 & 5.28E-10 & 1.74E-11 & 7.05E-12 & 5.25E-10 & 1.73E-11 & 3.58E-12 & 2.60E-10 & 9.62E-12 & 1.15E-11             \\
		& 3  & 5.71E-09 & 3.81E-07 & 1.28E-08 & 5.71E-09 & 3.81E-07 & 1.28E-08 & 5.16E-09 & 3.49E-07 & 1.19E-08 & 1.64E-08             \\
		& 5  & 6.64E-06 & 4.98E-04 & 2.53E-05 & 6.65E-06 & 4.98E-04 & 2.53E-05 & 6.56E-06 & 4.94E-04 & 2.52E-05 & 3.01E-05             \\
		& 10 & 1.92E-02 & 3.29E-01 & 2.94E-02 & 1.92E-02 & 3.29E-01 & 2.94E-02 & 1.92E-02 & 3.29E-01 & 2.94E-02 & 3.62E-02             \\
		& 30 & 2.17E+02 & 1.02E+03 & 5.55E+01 & 2.18E+02 & 1.02E+03 & 5.55E+01 & 2.18E+02 & 1.02E+03 & 5.55E+01 & 5.00E+01             \\
		&&&&&&&&&&&\\
$^{61}$Fe &1 & 1.00E-02 &    ---   & 6.47E-03 & 1.00E-02 &    ---   & 6.47E-03 & 9.98E-03 &    ---   & 6.43E-03 & 2.11E-02             \\
		& 2  & 2.30E-02 &    ---   & 1.04E-02 & 2.30E-02 &    ---   & 1.04E-02 & 2.29E-02 &    ---   & 1.04E-02 & 2.25E-02             \\
		& 3  & 3.68E-02 &    ---   & 1.51E-02 & 3.68E-02 &    ---   & 1.51E-02 & 3.67E-02 &    ---   & 1.51E-02 & 2.61E-02             \\
		& 5  & 8.04E-02 &    ---   & 5.78E-02 & 8.04E-02 &    ---   & 5.78E-02 & 8.02E-02 &    ---   & 5.77E-02 & 6.98E-02             \\
		& 10 & 5.51E-01 &    ---   & 6.95E-01 & 5.51E-01 &    ---   & 6.95E-01 & 5.50E-01 &    ---   & 6.95E-01 & 5.79E-01             \\
		& 30 & 3.10E+03 &    ---   & 1.31E+02 & 3.12E+03 &    ---   & 1.31E+02 & 3.12E+03 &    ---   & 1.31E+02 & 7.73E+01             \\
		&&&&&&&&&&&\\
$^{62}$Fe &1 & 4.08E-03 &    ---   & 1.27E-02 & 4.08E-03 &    ---   & 1.27E-02 & 4.03E-03 &    ---   & 1.25E-02 & 8.07E-03             \\
		& 2  & 4.10E-03 &    ---   & 1.27E-02 & 4.10E-03 &    ---   & 1.27E-02 & 4.06E-03 &    ---   & 1.26E-02 & 8.43E-03             \\
		& 3  & 4.38E-03 &    ---   & 1.50E-02 & 4.38E-03 &    ---   & 1.50E-02 & 4.33E-03 &    ---   & 1.48E-02 & 1.11E-02             \\
		& 5  & 1.67E-02 &    ---   & 7.83E-02 & 1.67E-02 &    ---   & 7.83E-02 & 1.66E-02 &    ---   & 7.82E-02 & 7.33E-02             \\
		& 10 & 1.23E+00 &    ---   & 1.43E+00 & 1.23E+00 &    ---   & 1.43E+00 & 1.23E+00 &    ---   & 1.43E+00 & 1.62E+00             \\
		& 30 & 7.74E+02 &    ---   & 1.45E+02 & 7.78E+02 &    ---   & 1.45E+02 & 7.78E+02 &    ---   & 1.45E+02 & 9.18E+01             \\
	&&&&&&&&&&&\\
$^{63}$Fe &1 & 1.67E+00 &    ---   & 2.55E-01 & 1.67E+00 &    ---   & 2.55E-01 & 1.67E+00 &    ---   & 2.55E-01 & ---    \\
		& 2  & 1.80E+00 &    ---   & 3.41E-01 & 1.80E+00 &    ---   & 3.41E-01 & 1.80E+00 &    ---   & 3.41E-01 &   ---  \\
		& 3  & 1.81E+00 &    ---   & 4.94E-01 & 1.81E+00 &    ---   & 4.94E-01 & 1.80E+00 &    ---   & 4.93E-01 &   ---  \\
		& 5  & 1.87E+00 &    ---   & 1.03E+00 & 1.87E+00 &    ---   & 1.03E+00 & 1.87E+00 &    ---   & 1.03E+00 &   ---  \\
		& 10 & 5.00E+00 &    ---   & 3.45E+00 & 5.01E+00 &    ---   & 3.45E+00 & 5.00E+00 &    ---   & 3.45E+00 &   ---  \\
		& 30 & 6.70E+03 &    ---   & 1.57E+02 & 6.71E+03 &    ---   & 1.57E+02 & 6.73E+03 &    ---   & 1.57E+02 &  ---  \\
			\hline
	\end{tabular}}
\end{table*}
\begin{table*}[]
	\scriptsize\caption{Same as Table~\ref{T1an} but for higher core densities.}
	\label{T2an}
	\centering
	\scalebox{.70}{
		\begin{tabular}{c|c|cccc|cccc|cccc|}
			\hline
			\multicolumn{7}{l}{} & \multicolumn{1}{l}{$\lambda^{\bar{\nu}}$ (MeV.$s^{-1}$)} & \multicolumn{6}{l}{}        \\
			\cline{1-14} \multicolumn{1}{l}{} &  & \multicolumn{4}{c}{$\rho$$\it Y_{e}$ = $10^7$ (g.cm$^{-3}$)}        & \multicolumn{4}{c}{$\rho$$\it Y_{e}$ = $10^9$ (g.cm$^{-3}$)}& \multicolumn{4}{c}{$\rho$$\it Y_{e}$ = $10^{11}$ (g.cm$^{-3}$)}         \\
			\cline{3-14} \multicolumn{1}{l}{Nuclei} &       $T_9$ (GK) & $\lambda_{pn-QRPA}^{\bar{\nu}}$& $\lambda_{IPM}^{\bar{\nu}}$    & $\lambda_{LSSM}^{\bar{\nu}}$   &$\lambda_{GXPF1J}^{\bar{\nu}}$&$\lambda_{pn-QRPA}^{\bar{\nu}}$& $\lambda_{IPM}^{\bar{\nu}}$    & $\lambda_{LSSM}^{\bar{\nu}}$    &$\lambda_{GXPF1J}^{\bar{\nu}}$    & $\lambda_{pn-QRPA}^{\bar{\nu}}$& $\lambda_{IPM}^{\bar{\nu}}$    & $\lambda_{LSSM}^{\bar{\nu}}$ &$\lambda_{GXPF1J}^{\bar{\nu}}$  \\
			\hline\\
$^{57}$Fe & 1& 1.52E-15 & 5.14E-15 & 1.90E-16 & 2.56E-16             & 1.51E-33 & 8.75E-32 & 3.55E-32 & 3.27E-31             &  $<$ 1.00E-100 & $<$1.00E-100 & $<$1.00E-100 & 1.43E-120            \\
		& 2  & 4.91E-10 & 1.52E-09 & 1.38E-10 & 1.04E-10             & 1.42E-18 & 1.19E-16 & 3.30E-17 & 9.64E-17             & 7.18E-66  & 3.59E-64  & 3.66E-64  & 1.30E-61             \\
		& 3  & 8.05E-08 & 1.10E-06 & 4.88E-08 & 3.91E-08             & 2.22E-13 & 1.42E-11 & 5.89E-12 & 1.21E-11             & 6.10E-45  & 2.92E-43  & 3.27E-43  & 1.70E-41             \\
		& 5  & 1.99E-05 & 3.44E-04 & 2.85E-05 & 2.33E-05             & 4.82E-09 & 2.19E-07 & 1.64E-07 & 2.57E-07             & 4.86E-28  & 2.41E-26  & 3.19E-26  & 6.68E-25             \\
		& 10 & 1.53E-02 & 1.17E-01 & 1.44E-02 & 1.15E-02             & 9.29E-05 & 1.53E-03 & 7.33E-04 & 8.17E-04             & 2.18E-14  & 3.48E-13  & 3.58E-13  & 3.15E-12             \\
		& 30 & 1.03E+03 & 6.00E+02 & 3.99E+01 & 2.97E+01             & 4.86E+02 & 2.86E+02 & 1.91E+01 & 1.51E+01             & 1.42E-01  & 8.45E-02  & 5.98E-03  & 6.84E-03             \\
		&&&&&&&&&&&\\
$^{58}$Fe & 1& 2.53E-23 & 2.66E-21 & 4.35E-22 & 7.73E-22             & 3.07E-40 & 5.87E-38 & 1.75E-38 & 3.38E-37             & $<$1.00E-100 & $<$1.00E-100 & $<$1.00E-100 & 3.11E-126            \\
		& 2  & 1.63E-13 & 6.31E-12 & 7.29E-13 & 1.46E-12             & 1.85E-21 & 4.05E-19 & 1.47E-19 & 5.56E-19             & 1.25E-68  & 2.08E-66  & 1.83E-66  & 1.22E-63             \\
		& 3  & 9.75E-10 & 1.03E-07 & 4.05E-09 & 5.00E-09             & 6.27E-15 & 1.31E-12 & 5.64E-13 & 1.42E-12             & 2.35E-46  & 3.65E-44  & 3.52E-44  & 2.75E-42             \\
		& 5  & 2.89E-06 & 3.39E-04 & 1.90E-05 & 1.99E-05             & 2.06E-09 & 2.72E-07 & 1.49E-07 & 3.17E-07             & 2.84E-28  & 2.83E-26  & 3.15E-26  & 1.06E-24             \\
		& 10 & 1.56E-02 & 2.75E-01 & 2.64E-02 & 3.33E-02             & 1.44E-04 & 3.77E-03 & 1.93E-03 & 3.50E-03             & 3.91E-14  & 8.85E-13  & 9.59E-13  & 1.45E-11             \\
		& 30 & 2.16E+02 & 1.02E+03 & 5.50E+01 & 5.01E+01             & 1.02E+02 & 4.84E+02 & 2.65E+01 & 2.58E+01             & 2.98E-02  & 1.43E-01  & 8.77E-03  & 1.32E-02             \\
			&&&&&&&&&&&\\
$^{61}$Fe &1 & 7.48E-03 &   ---    & 4.46E-03 & 1.70E-02             & 1.87E-09 &   ---    & 7.45E-10 & 4.17E-08             & $<$1.00E-100 &   ---     & $<$1.00E-100 & 1.83E-95             \\
		& 2  & 1.89E-02 &   ---    & 8.00E-03 & 1.90E-02             & 3.04E-06 &   ---    & 1.26E-06 & 5.86E-06             & 4.52E-53  &   ---     & 6.89E-52  & 4.37E-49             \\
		& 3  & 3.16E-02 &   ---    & 1.24E-02 & 2.30E-02             & 6.76E-05 &   ---    & 6.71E-05 & 1.64E-04             & 3.63E-36  &   ---     & 5.51E-35  & 3.69E-33             \\
		& 5  & 7.16E-02 &   ---    & 5.22E-02 & 6.47E-02             & 1.29E-03 &   ---    & 3.41E-03 & 5.30E-03             & 2.01E-22  &   ---     & 2.97E-21  & 9.98E-20             \\
		& 10 & 4.90E-01 &   ---    & 6.55E-01 & 5.56E-01             & 3.36E-02 &   ---    & 1.28E-01 & 1.28E-01             & 1.18E-11  &   ---     & 1.16E-10  & 8.45E-10             \\
		& 30 & 3.10E+03 &   ---    & 1.30E+02 & 7.74E+01             & 1.47E+03 &   ---    & 6.24E+01 & 3.98E+01             & 4.33E-01  &   ---     & 2.11E-02  & 2.08E-02             \\
		&&&&&&&&&&&\\
$^{62}$Fe & 1& 1.95E-03 &   ---    & 5.74E-03 & 4.10E-03             & 4.85E-18 &   ---    & 7.71E-16 & 2.01E-14             & $<$1.00E-100 &   ---     & $<$1.00E-100 & 2.72E-102            \\
		& 2  & 2.20E-03 &   ---    & 6.67E-03 & 4.99E-03             & 3.98E-10 &   ---    & 1.48E-08 & 2.93E-08             & 2.92E-55  &   ---     & 2.14E-54  & 7.46E-52             \\
		& 3  & 2.52E-03 &   ---    & 9.53E-03 & 7.89E-03             & 6.28E-07 &   ---    & 9.93E-06 & 1.39E-05             & 5.19E-37  &   ---     & 3.62E-36  & 1.66E-34             \\
		& 5  & 1.17E-02 &   ---    & 6.78E-02 & 6.56E-02             & 4.69E-04 &   ---    & 2.96E-03 & 4.42E-03             & 4.41E-22  &   ---     & 1.79E-21  & 6.50E-20             \\
		& 10 & 1.13E+00 &   ---    & 1.35E+00 & 1.58E+00             & 1.63E-01 &   ---    & 2.45E-01 & 3.85E-01             & 1.69E-10  &   ---     & 1.91E-10  & 2.31E-09             \\
		& 30 & 7.73E+02 &   ---    & 1.43E+02 & 9.20E+01             & 3.67E+02 &   ---    & 7.00E+01 & 4.86E+01             & 1.12E-01  &   ---     & 2.55E-02  & 3.29E-02             \\
			&&&&&&&&&&&\\
$^{63}$Fe& 1 & 1.55E+00 &   ---    & 2.17E-01 &  ---			     & 4.67E-02 &   ---    & 3.04E-04 &  	---				 & 7.23E-92  &   ---     & 2.81E-91  & ---   \\
		& 2  & 1.67E+00 &   ---    & 2.99E-01 &---			         & 5.82E-02 &   ---    & 3.70E-03 &    ---               & 1.96E-47  &   ---     & 1.80E-46  &   --- \\
		& 3  & 1.69E+00 &   ---    & 4.48E-01 &   --- 				 & 7.23E-02 &   ---    & 1.88E-02 & ---   			     & 2.26E-32  &   ---     & 2.34E-31  &    ---\\
		& 5  & 1.76E+00 &   ---    & 9.75E-01 &  	---			     & 1.22E-01 &   ---    & 1.25E-01 &    			---		 & 4.58E-20  &   ---     & 4.52E-19  &    ---\\
		& 10 & 4.63E+00 &   ---    & 3.32E+00 & 		---		     & 6.17E-01 &   ---    & 9.44E-01 &    		---			 & 2.95E-10  &   ---     & 1.39E-09  &    ---\\
		& 30 & 6.67E+03 &   ---    & 1.56E+02 & 			---	     & 3.17E+03 &   ---    & 7.60E+01 &    				---	 & 9.44E-01  &   ---     & 2.74E-02  &    ---\\
			\hline
	\end{tabular}}
\end{table*}


\end{document}